\documentclass[a4paper,11pt]{article}
\usepackage{jheppub} 
\usepackage{lineno}

\usepackage[utf8]{inputenc}
\usepackage[colorlinks=true,citecolor=blue,linkcolor=blue]{hyperref}
\usepackage[normalem]{ulem}
\usepackage{amsmath,amssymb}
\usepackage{epsfig}
\usepackage{graphicx}               
\usepackage{url}
\usepackage{color}
\usepackage{slashed}
\usepackage{multirow}
\usepackage{placeins}
\usepackage[dvipsnames]{xcolor}
\usepackage{bm}
\usepackage{epstopdf}
\usepackage{soul}
\usepackage{tikz}
\usepackage[capitalise, english]{cleveref}
\usepackage{siunitx}
\usepackage{xspace}
\usetikzlibrary{trees}
\usetikzlibrary{decorations.pathmorphing}
\usetikzlibrary{decorations.markings}

\newcommand\myshade{80}
\colorlet{mylinkcolor}{ForestGreen}
\colorlet{mycitecolor}{Red}
\colorlet{myurlcolor}{violet}

\hypersetup{
  linkcolor  = mylinkcolor!\myshade!black,
  citecolor  = mycitecolor!\myshade!black,
  urlcolor   = myurlcolor!\myshade!black,
  colorlinks = true
}

\definecolor{jblue}{RGB}{20,50,100}
\definecolor{npurple}{RGB} {153, 51, 204}
\definecolor{wred}{RGB}{217,0,56}
\definecolor{white}{RGB}{255,255,255}

\definecolor{korange}{RGB}{235, 80,  43}
\definecolor{korange2}{RGB}{245, 100,  63}
\definecolor{kyelloworange}{RGB}{255, 210,  110}
\definecolor{kyelloworange2}{RGB}{240, 170,  90}
\definecolor{kred}{RGB}{204,  102, 153}
\definecolor{kpurple}{RGB}{153,  61, 190}
\definecolor{kpurplelight}{RGB}{213,  161, 230}

\allowdisplaybreaks

\newcommand{\vq}{\langle \Phi_q \rangle}
\newcommand{\vh}{\langle \Phi_H \rangle}

\makeatletter
\gdef\@fpheader{}
\makeatother
\begin{document}

\title{Deconstructed Hypercharge: \\
A Natural Model of Flavour}

\author{Joe Davighi}
\emailAdd{joe.davighi@physik.uzh.ch}

\author{and Ben A. Stefanek}
\emailAdd{ben.stefanek@physik.uzh.ch}

\affiliation{Physik-Institut, Universit\"at Z\"urich, CH-8057 Z\"urich, Switzerland}

\date{\today}

\preprint{ZU-TH 24/23}

\abstract{
The flavour puzzle is one of the greatest mysteries in particle physics. A `flavour deconstruction' of the electroweak gauge symmetry, by promoting at least part of it to the product of a third family factor (under which the Higgs is charged) times a light family factor, allows one to address the flavour puzzle at a low scale due to accidentally realised $U(2)^5$ flavour symmetries. 
The unavoidable consequence is new heavy gauge bosons with direct couplings to the Higgs, threatening the stability of the electroweak scale. In this work, we propose a UV complete model of flavour based on deconstructing only hypercharge. We find that the model satisfies finite naturalness criteria, benefiting from the smallness of the hypercharge gauge coupling in controlling radiative Higgs mass corrections and passing phenomenological bounds. Our setup allows one to begin explaining flavour at the TeV scale, while dynamics solving the large hierarchy problem can lie at a higher scale up to around 10 TeV - without worsening the unavoidable little hierarchy problem. The low-energy phenomenology of the model is dominated by a single $Z'$ gauge boson with chiral and flavour non-universal couplings, with mass as light as a few TeV thanks to the $U(2)^5$ symmetry. The natural parameter space of the model will be probed by the HL-LHC and unavoidably leads to large positive shifts in the $W$-boson mass, as well as an enhancement in $\mathcal{B}(B_{s,d} \to \mu^+ \mu^-)$. Finally, we show that a future electroweak precision machine such as FCC-ee easily has the reach to fully exclude the model.
}

\maketitle

\section{Introduction}
Flavour is a rich source of structure in the Standard Model (SM). The huge hierarchies in quark and charged lepton masses, which span six orders of magnitude, together with the hierarchy of quark mixing angles, go unexplained in the SM. This constitutes the {\em SM flavour puzzle}, which begs for an explanation in the form of dynamics {\em beyond} the SM (BSM).

On the flip side of the coin, flavour has a crucial impact on the search for BSM physics -- whether or not the BSM particles explain the SM flavour puzzle -- in two general ways.
Firstly, as has long been known, indirect bounds on flavour-changing neutral currents (FCNC) push the scale of flavour anarchic new physics (NP) well above the PeV scale~\cite{Silvestrini:2018dos,EuropeanStrategyforParticlePhysicsPreparatoryGroup:2019qin}. The fact that flavour points to the PeV scale, while there are compelling arguments to expect BSM physics much closer to the electroweak scale, is often referred to as the {\em BSM flavour puzzle.} It follows that any NP within direct reach of current or near future particle experiments \emph{must} have a very particular flavour structure in order to pass these bounds. A key example of such flavour protection is the Minimal Flavour Violation (MFV) paradigm~\cite{DAmbrosio:2002vsn}, which lowers the scale of NP with tree-level FCNC to the 10 TeV ballpark. To reach the TeV scale, one generally requires additional suppression, {\em e.g.} via loops.
Secondly, past and present particle colliders have used beams composed (predominantly) of first generation fermions, namely (anti-) electrons and (anti-) protons, meaning that NP coupled dominantly to the heavy flavours is much less constrained by collider bounds -- see {\em e.g.}~\cite{Allwicher:2022gkm,Greljo:2022jac} for recent studies. 
The current state of play is that direct collider searches at high-$p_T$ exclude most NP models based on flavour-universal $U(3)^n$ symmetries up to scales also $\mathcal{O}(10)$ TeV, due to their couplings to valence quarks.\footnote{Other low-energy experiments, including neutrino scattering measurements as well as probes of atomic parity violation (APV), provide complementary constraints on $U(3)$-invariant new physics, with the bounds being driven by the large couplings to the first generation. See Ref.~\cite{Greljo:2023adz} (also~\cite{Falkowski:2017pss,Breso-Pla:2023tnz}) for a comprehensive analysis of $U(3)$-invariant NP including these datasets.}
However, there remains plenty of wiggle room to discover TeV scale NP particles if they couple mostly to the third generation.

Such a dominant coupling of NP to the third {\em vs.} light generations can be described via approximate $U(2)^5 \equiv U(2)_q \times U(2)_u \times U(2)_d \times U(2)_\ell \times U(2)_e$ global flavour symmetries (a subgroup of the $U(3)^5$ symmetry mentioned above) under which the light generations transform as doublets and the third generation as singlets.
Via minimal breaking patterns, these same $U(2)^5$ symmetries can efficiently account for the observed fermion mass and mixing hierarchies, while simultaneously suppressing flavour violation in the light families~\cite{Barbieri:2011ci,Isidori:2012ts,Barbieri:2012uh,Fuentes-Martin:2019mun,Kagan:2009bn} at least as well as in MFV, as is needed to also pass the stringent flavour bounds. Appropriately broken $U(2)^5$ flavour symmetries therefore allow for phenomenologically viable NP particles with masses $\sim 1$ TeV, where the improvement over $U(3)^n$ comes from the freedom to suppress NP couplings to the light generations. 
Putting things together, BSM dynamics with accidental approximate $U(2)$ symmetries can offer solutions to the SM flavour puzzle without introducing a BSM flavour puzzle, all while being as light as a few TeV.

These $U(2)^5$ symmetries can themselves be realised accidentally via flavour non-universal gauge symmetries. This idea originated with the study of completely `horizontal' non-universal gauge symmetries~\cite{Froggatt:1978nt}, {\em i.e.} those that commute with the SM gauge symmetry, which were typically (but not always~\cite{Grinstein:2010ve}) imagined to be broken at a very high scale. The idea has been re-invigorated in recent years~\cite{Bordone:2017bld,Greljo:2018tuh,Fuentes-Martin:2020pww,Fuentes-Martin:2020bnh,Davighi:2022fer,Fuentes-Martin:2022xnb,FernandezNavarro:2022gst,Davighi:2022bqf,Stefanek:2022doo} through a more intricate setup in which 
one `flavour deconstructs' various factors of the SM gauge symmetry, by promoting that factor to the product of a light generation part and a third generation part, thereby delivering $U(2)^5$ accidental symmetry.  We emphasize that here (in contrast to the horizontal case) the flavour non-universality is intrinsic in the UV embedding of the SM gauge interactions themselves, with universality emerging as a low-energy accident.

The various options for deconstructing the SM gauge symmetry were explored recently in Ref.~\cite{Davighi:2023iks}.
For example, by deconstructing only colour, namely $SU(3)_c \rightarrow SU(3)_3 \times SU(3)_{12}$, one obtains $U(2)^3$ quark flavour symmetries that can explain the smallness of the CKM elements $|V_{cb,ub}|\ll 1$. However, one cannot explain the mass hierarchies $m_2 \ll m_3$, simply because the Higgs is colourless. To simultaneously explain the smallness of $|V_{cb,ub}|$ and $m_2 \ll m_3$, one must deconstruct at least part of the electroweak (EW) gauge symmetry $SU(2)_L \times U(1)_Y$, and charge the Higgs under the third family factor. The Higgs vacuum expectation value (VEV) then picks out a particular direction in flavour space, giving mass only to the third family at leading order.  
In the present work we examine in detail what we consider to be the most minimal option from the bottom-up, namely to deconstruct SM hypercharge, which delivers the full $U(2)^5$ symmetry while extending the SM gauge sector by only a single extra gauge boson.\footnote{Deconstructing only $SU(2)_L$ can also explain why $m_2 \ll m_3$. Although this structure would in principle allow the full third row of the Yukawa matrices to be populated, right-handed rotations remain unphysical in this model (as in the SM) as the flavour symmetry without Yukawas is $U(2)_q \times U(2)_l \times U(3)_u \times U(3)_d \times U(3)_e$.} 

In all such models with a flavour-deconstructed EW symmetry, there are new heavy gauge bosons that couple directly to the Higgs. This unavoidably leads to tree-level NP corrections to electroweak precision observables (EWPO), as well as 1-loop contributions to the Higgs mass-squared parameter $\delta m_h^2$, which threatens to destabilize the EW scale~\cite{Farina:2013mla}. However, radiative Higgs mass corrections are minimized when the gauge boson masses are as low as is phenomenologically viable -- which recall is not much more than 1 TeV when invoking $U(2)^n$ flavour symmetries.
In this way, we see that solving the flavour puzzle at a low scale is closely linked to the (in)stability of the EW scale in this intermediate energy r\'egime~\cite{Davighi:2023iks}, where it appears to be a fact of life that supersymmetry (or compositeness of the Higgs, or whatever else solves the `large hierarchy problem') does not yet operate to protect $m_h^2$.
If we are fortunate enough that the flavour puzzle is resolved at low scales, and if we do not want to give up hope on $m_h^2$ being fundamentally calculable (and not fine-tuned beyond the first few digits), then it is important to identify which viable options are most natural.

A particularly promising direction is the deconstructed hypercharge (DH) option, where one benefits from $g_Y$ being the smallest SM gauge coupling, both in suppressing tree-level corrections to EWPO $\propto g_Y^2 /m_X^2$ as well as radiative Higgs mass corrections from the gauge sector scaling as
$\delta m_h^2 \propto g_Y^2 m_X^2/(4\pi)^2$. As could already be anticipated, the reciprocal scaling of these corrections with $m_X$ prefers the gauge field to have an intermediate mass of a few TeV, where both constraints (as well as high-$p_T$ bounds) can be simultaneously satisfied.

In this paper we develop a fully renormalizable model of flavour based on the DH gauge symmetry, using naturalness as a guiding principle. In particular, following the philosophy of `finite naturalness'~\cite{Farina:2013mla}, we demand that the calculable finite loop corrections to $m_h^2$ are not greater than $(1\text{~TeV})^2$, which is the scale of the little hierarchy problem indicated by current experimental data.
The low-energy phenomenology of our DH model is dominated by a single $Z'$ gauge boson, which has highly chiral and flavour non-universal couplings to the SM fermions.
We remark that the symmetry breaking structure is in fact equivalent to that of the ``third family hypercharge ($Y_3$) model", introduced by Allanach and one of us in~\cite{Allanach:2018lvl} and developed in~\cite{Allanach:2019iiy,Allanach:2021bbd,Davighi:2021oel,Allanach:2021gmj,Allanach:2021kzj,Allanach:2022bik}, up to a linear field redefinition and a compensating (large) shift of the kinetic mixing angle~\cite{Greljo:2021npi}.\footnote{The $Y_3$ model was originally introduced
as a possible $Z'$ explanation of the LFUV observed in the $b\to s \ell\ell$ system, for which the evidence has gone away with recent LHCb data~\cite{LHCb:2022qnv} (even though discrepancies in $b\to s\mu\mu$ transitions remain). Various other $Z'$ models with $U(2)^n$ accidental symmetries have also been studied, based on different charge assignments, for example in Refs.~\cite{Falkowski:2015zwa,Calibbi:2019lvs,Alguero:2022est}.
} 
The origin of the light Yukawas in these phenomenological $Z'$ models was never studied, and for the purposes of explaining flavour the DH ``basis choice" (with zero kinetic mixing) is more natural.

To build towards a UV complete model, we envisage that the first layer of NP, namely the $Z'$ associated with the deconstructed gauge symmetry as well as the scalar sector that breaks it to the SM, lies at a low scale $f \approx$ TeV. This is required by finite naturalness, to avoid a finely-tuned electroweak scale irrespective of what BSM dynamics resides at even heavier scales. The light Yukawas are realised as higher-dimensional operators at this scale, which come from integrating out heavier states at a scale $4\pi f \approx$ 10 TeV. At this second threshold, we also envisage a solution to the large hierarchy problem such as supersymmetry (SUSY) or compositeness, which protects the scale $f$ from radiative corrections. While we do not commit to any particular UV dynamics at the scale $4\pi f$ in this paper (meaning we do not explicitly specify how the {\em large} hierarchy problem is solved), we do take care to consider the impact of such a sector on sensitive flavour observables via a model-independent spurion analysis. As we will show, achieving consistency with flavour bounds in the case where the UV theory is strongly coupled severely restricts how the $U(2)^5$ flavour symmetry may be broken. 
Indeed, without having to specify an explicit solution to the large hierarchy problem, we are already pointed to a very specific UV completion of the Yukawa sector of the DH model via generic spurion arguments.
In particular, the CKM elements $V_{cb}$ and $V_{ub}$ arise from integrating out a heavy vector-like quark doublet, while the light fermion masses and Cabibbo mixing angle originate instead from a heavy Higgs doublet -- with both of these new states residing at the second threshold $\sim 4 \pi f$. We emphasize that $f$ should be taken as low as is phenomenologically viable in order to minimize the unavoidable `little hierarchy problem', {\em i.e.} the `tree-level' $v^2/f^2$ sized fine-tuning one must swallow such that the Higgs mass resides at the EW scale rather than its natural scale $f$.

The natural parameter space of the DH model has a
rich and predictive phenomenology that is being probed at the LHC and in precision flavour experiments. For example:
\begin{itemize}
    \item If the model fully explains the hierarchy $m_c/m_t$ via a ratio of scales, given also our naturalness constraints, then the $Z'$ is predicted to be rather light. Since it couples directly to the Higgs, one cannot avoid large effects in EWPOs, the largest being a positive shift in the $W$-boson mass and corrections of the SM $Z$-boson couplings to right-handed charged leptons. The present status of $M_W$ measurements accommodates such a sizeable positive shift, such that the model improves the EW fit if the $Z'$ is relatively light, with $Z\rightarrow e_R\,  e_R$ ($e = e,\mu,\tau$) providing the dominant constraint. Finally, we show that the FCC-ee will easily have enough reach to probe the entire (natural) parameter space of our DH model.
    \item ATLAS and CMS searches at high-$p_T$, in all di-lepton final states (possibly with $b$-tagging), are probing the natural parameter space of the model. Additionally, we give a projection for how the high-luminosity LHC is expected to improve these bounds.
    \item Naturalness together with flavour constraints require that the CKM matrix originates mainly from the down-quark sector. This leads to important effects in $b\to s\ell\ell$ transitions, either via direct couplings to light leptons or large $Z-Z'$ mixing, that cannot be decoupled in the natural parameter space of the model. These effects are flavour universal in $e$ vs $\mu$, and in the case of large $Z-Z'$ mixing they yield $|C_{10}^{\mu,e}| \gg |C_9^{\mu,e}|$, resulting in a significant enhancement of $\mathcal{B}(B_{s,d} \to \mu^+ \mu^-)$.
\end{itemize}
This paper is organized as follows. In \S~\ref{sec:model} we introduce the basics of the deconstructed hypercharge model, as a TeV scale effective field theory (EFT) that delivers a $U(2)^5$ flavour symmetry. In \S~\ref{sec:UV-completion} we set out a UV completion of the model that we suggest is most natural. We quantify the fine-tuning of $m_h^2$ in \S~\ref{sec:naturalness}. We analyse the phenomenology of the model in \S~\ref{sec:pheno}, before concluding. 

{\em Note added}: We note that while finalising this manuscript, Ref.~\cite{FernandezNavarro:2023rhv} was published on the \texttt{arXiv} presenting a `tri-hypercharge' model that has some overlap with the present work.

\section{Model Basics and EFT}
\label{sec:model}
\begin{table}[ht]
\renewcommand{\arraystretch}{1.1}
\begin{center}
\begin{tabular}{|c|c|c|c|c|}
\hline 
 {\rm Field } & $SU(3)_c$    &   $SU(2)_L$   & $U(1)_3$  & $U(1)_{12}$ \\ \hline
 $q_{L}^{i}$  & {\bf 3}  &  {\bf 2}  &  0  & 1/6      \\ 
 $u_{R}^{i}$  & {\bf 3}  &  {\bf 1}  &  0  & 2/3     \\ 
 $d_{R}^{i}$  & {\bf 3}  &  {\bf 1}  &  0  & -1/3      \\ 
 $\ell_{L}^{i}$  & {\bf 1}  &  {\bf 2}  & 0  & -1/2     \\ 
 $e_{R}^{i}$  & {\bf 1}  &  {\bf 1}  &  0  & -1    \\ 
  \hline 
   \hline 
 $q_{L}^{3}$  & {\bf 3}  &  {\bf 2}  &   1/6  & 0    \\ 
 $t_{R}$  & {\bf 3}  &  {\bf 1}  &   2/3 & 0     \\ 
 $b_{R}$  & {\bf 3}  &  {\bf 1}  &  -1/3 & 0    \\ 
 $\ell_{L}^{3}$  & {\bf 1}  &  {\bf 2}  & -1/2 & 0     \\ 
 $\tau_{R}$  & {\bf 1}  &  {\bf 1}  &   -1 & 0   \\ 
 \hline
   \hline 
    $H_3$  & {\bf 1}  &  {\bf 2}  &   1/2  & 0    \\ 
    $\Phi_H$  & {\bf 1}  &  {\bf 1}  &  -1/2 & 1/2    \\ 
 $\Phi_q$  & {\bf 1}  &  {\bf 1}  &   -1/6 & 1/6     \\ 
  \hline
\end{tabular}
\end{center}
\caption{Matter content of the model in $\mathcal{G}_{\rm DH}$-symmetric phase. The flavour index $i = 1,2$ runs over the 1st and 2nd family fermions.
}
\label{tab:fieldcontent}
\end{table}

\subsection{Gauge sector and symmetry breaking}

We consider a model of flavour based on the deconstructed hypercharge (DH) gauge group,
\begin{equation}
\mathcal{G}_{\rm DH} = SU(3)_c \times SU(2)_L \times U(1)_3 \times U(1)_{12} \,,
\end{equation}
with gauge couplings $g_s, g_L, g_3$ and $g_{12}$, respectively. The light family fermions 
are charged under the light family hypercharge group $U(1)_{12}$ while the third family and $H_3$ are charged under the third family hypercharge factor $U(1)_3$ as shown in~\cref{tab:fieldcontent}. Since anomaly cancellation occurs family-by-family within the SM, this setup is manifestly free of gauge anomalies (nor are there non-perturbative gauge anomalies~\cite{Davighi:2019rcd}).  The $\mathcal{G}_{\rm DH}$ gauge symmetry allows only the third family Yukawa couplings with $H_3$, realizing an accidental $U(2)^5 = U(2)_q \times U(2)_u \times U(2)_d \times U(2)_\ell \times U(2)_e$ global flavour symmetry.

The $U(1)_{3} \times U(1)_{12}$ part of $\mathcal{G}_{\rm DH}$ is spontaneously broken to the diagonal SM hypercharge group $U(1)_Y$ by the VEVs of two complex scalar fields $\langle \Phi_{H,q}\rangle$, whose charges are recorded in Table~\ref{tab:fieldcontent}. While one scalar is sufficient to realise the symmetry breaking pattern, we posit a second scalar that plays a key role in generating the light SM fermion Yukawa couplings in this model -- see \S \ref{sec:light-yuk-eft} and \ref{sec:UV-completion}. We assume that there is negligible kinetic mixing between the two $U(1)$ factors at the energy scales of relevance, as would be the case, for example, if they emerged from some semi-simple gauge symmetry at a higher energy scale.\footnote{
The kinetic mixing parameter $\epsilon$, defined via $\mathcal{L}\supset \frac{1}{2}\epsilon F_{12}^{\mu\nu} F_{3,\mu\nu}$, is generated at 1-loop due to the scalar link fields $\Phi_{H,q}$ that are charged under both $U(1)$ groups. We have $d\epsilon/d\log\mu=-g_{12} g_3 (X_H^2 + X_q^2)/12\pi^2$ (see {\em e.g.}~\cite{Greljo:2021npi}), where $X_{H(q)}$ denotes the SM hypercharge of the Higgs (left-handed quark doublet). If $\epsilon$ is zero at some high unification scale $\Lambda_{\rm{GUT}}$, then it runs slowly; for illustration, we numerically find that $\epsilon$ remains at the few percent level even running from a scale $\Lambda_{\rm{GUT}} \sim 10^{13}$ TeV down to 10 TeV.
}

One linear combination of the original gauge fields remains massless and is identified with the SM hypercharge gauge boson $B_{\mu}$, while the orthogonal combination is a massive $Z'_{\mu}$ vector boson. Its mass squared is
\begin{equation} \label{eq:Zp-mass}
M_{Z'}^2 = \frac{2g_Y^2}{\sin^2 2\theta} \left[ \langle \Phi_H \rangle^2 + \frac{1}{9} \langle \Phi_q \rangle^2 \right] \,,
\end{equation}
where 
\begin{equation}
    \tan \theta := g_{12} / g_3
\end{equation} 
is the gauge mixing angle, analogous to the Weinberg angle of the electroweak theory. Since we must identify the unbroken $U(1)$ with (flavour-universal) SM hypercharge, with gauge coupling $g_Y$, the original two gauge couplings $g_{12}$ and $g_3$ may be traded for $g_Y$ and the mixing angle: 
\begin{equation}
g_Y = g_3 \sin\theta = g_{12} \cos\theta \,.
\end{equation}
On the other hand, the $Z'$ couplings are flavour non-universal. In the fermion gauge eigenbasis, its couplings are $\mathcal{L} \supset g_{Z'}^{ij} X_\psi \overline{\psi}_i \gamma^\mu Z^{\prime}_\mu \psi_j$, where the matrix $g_{Z^\prime}^{ij}$ is defined as
\begin{equation} \label{eq:Zp-couplings}
g_{Z^\prime}^{ij} = g_Y \, {\rm diag}(-\tan\theta, -\tan\theta, \,\cot\theta )\,,
\end{equation}
and where $X_\psi$ denotes the hypercharge of a SM fermion species $\psi \in \{q_L, u_R, d_R, \ell_L, e_R \}$.

\subsection{EFT for light Yukawas} \label{sec:light-yuk-eft}

Given the field content in Table~\ref{tab:fieldcontent}, 
the only renormalisable Yukawa couplings admitted are those for the third family:
\begin{align}
\mathcal{L}_{\rm Yuk} = -y^3_u \bar q_L^3 \tilde{H}_3 t_R -y^3_d \bar q_L^3 H_3 b_R  -y^3_e \bar \ell_L^3 H_3 \tau_R\, ,
\end{align}
as in the $Y_3$ model of~\cite{Allanach:2018lvl} and we define $\tilde{H}_3 = i\sigma_{2} H_{3}^*$.
One can write down higher-dimensional effective operators that would match onto Yukawa couplings for the light generations after $\Phi_{H,q}$ condense. At dimension-5, one expects left-handed mixing between light and heavy quarks to be generated:
\begin{equation}
\mathcal{L}_{d=5} \supset \frac{C_t^i}{\Lambda_q} \bar q_L^i \Phi_q \tilde{H}_3 t_R + \frac{C_b^i}{\Lambda_q} \bar q_L^i \Phi_q H_3 b_R \,,
\label{eq:LHd5}
\end{equation}
which will generate the CKM mixing angle $V_{cb}$ with a na\"ive suppression by a factor $\langle \Phi_q \rangle/\Lambda_q$. Similarly, one can have charged-lepton mixing via $\Phi_H$
\begin{equation} \label{eq:mu-tau}
\mathcal{L}_{d=5} \supset \frac{C_\tau^i}{\Lambda_\ell} \bar \ell_L^i \Phi_H^* H_{3} \tau_R \,.
\end{equation}
The right-handed mixing, on the other hand, is generated at dimension-6 via the operators:
\begin{equation}
\mathcal{L}_{d=6} \supset \frac{\kappa_u^i}{\Lambda_q \Lambda_H} \bar q_L^3 \Phi_q^* \Phi_H^* \tilde{H}_3 u_R^i + \frac{\kappa_d^i}{\Lambda_q \Lambda_H} \bar q_L^3 \Phi_q^* \Phi_H H_3 d_R^i \,,
\label{eq:d6RH}
\end{equation}
and so is naturally suppressed with respect to the left-handed mixing -- as is phemenologically required for a low-scale model.

The `light-light' Yukawa couplings, for both types of quarks and for charged leptons, are also expected to be generated at dimension-5:
\begin{equation}
\mathcal{L}_{d=5} \supset \frac{C_u^{ij}}{\Lambda_H} \bar q_L^i \Phi_H^* \tilde{H}_3 u_R^j + \frac{C_d^{ij}}{\Lambda_H} \bar q_L^i \Phi_H H_3 d_R^j +
\frac{C_e^{ij}}{\Lambda_H} \bar \ell_L^i \Phi_H H_3 e_R^j\,.
\label{eq:LHd5light}
\end{equation}
Thus, at the na\"ive level of counting operator dimensions, we expect the Yukawa matrices to have the hierarchical structures, which we write in `2+1' block form:
\begin{align}
    y_{u,d}\sim \begin{pmatrix} \frac{\langle\Phi_H \rangle}{\Lambda_H} & \frac{\langle\Phi_q \rangle}{\Lambda_q} \\
    \frac{\langle\Phi_H \rangle\langle\Phi_q \rangle}{\Lambda_H \Lambda_q} & 1
    \end{pmatrix}\, ,
    \quad\quad\quad
    y_{e}\sim \begin{pmatrix} \frac{\langle\Phi_H \rangle}{\Lambda_H} & \frac{\langle\Phi_H \rangle}{\Lambda_\ell} \\
    \frac{\langle\Phi_H \rangle\langle\Phi_q \rangle}{\Lambda_H \Lambda_\ell} & 1
    \end{pmatrix}\, .
\end{align}
These $U(2)$-based textures offer a promising starting point for explaining the fermion mass and mixing angle hierarchies with TeV scale NP (see {\em e.g.}~\cite{Fuentes-Martin:2019mun}).

One can also write down higher-dimensional Weinberg operators in our DH EFT, which give mass to the light neutrinos. These operators could have a different UV origin to that of the light Yukawas (which will be discussed in \S~\ref{sec:UV-completion}), and so can be suppressed by {\em a priori} independent effective NP scales. See Appendix~\ref{app:neutrino} for a discussion of one possible mechanism for neutrino mass generation in our model.

\section{UV Completing the Yukawa sector} \label{sec:UV-completion}

In this Section we aim to explicitly UV complete the deconstructed hypercharge model, at least from the point of view of its flavour structure. In other words, we will suggest the best-motivated origin for the effective operators detailed in \S~\ref{sec:light-yuk-eft}, taking into account both flavour physics constraints and requirements of naturalness.
Essentially, one needs extra fields that connect the symmetry breaking scalars $\Phi_{q,H}$ to the $H_3$ Higgs field and the (light) SM fermions. The most obvious options are for the extra fields to be either {\em vector-like fermions} (VLFs) or {\em extra scalars} with Higgs-like quantum numbers. 

The model that we focus on features two extra states (one VLF, $Q_{L,R}$, and one scalar, $H_{12}$), whose quantum numbers are written in Table~\ref{tab:fieldcontent2}. This choice effectively delivers the minimal $U(2)^5$ breaking pattern studied in Ref.~\cite{Fuentes-Martin:2019mun}, where $U(2)^5$ is broken only via spurions in the same representations as those needed to generate Yukawas within the SM. This makes it a particularly safe choice for passing flavour bounds~\cite{Barbieri:2011ci,Isidori:2012ts,Barbieri:2012uh}.

\begin{table}[ht]
\renewcommand{\arraystretch}{1.1}
\begin{center}
\begin{tabular}{|c|c|c|c|c|c|}
\hline 
 {\rm Field } & $SU(3)_c$    &   $SU(2)_L$   & $U(1)_3$  & $U(1)_{12}$ & {\rm Generates:} \\ \hline
   \hline 
    $H_{12}$  & {\bf 1}  &  {\bf 2}  &   0  & 1/2 & $y_{c,s,\mu,u,d,e}$, $V_{us}$   \\ 
    $Q_{L,R}$ & {\bf 3}  &  {\bf 2}  &   1/6  & 0  & $V_{cb}$, $V_{ub}$ \\
  \hline
\end{tabular}
\end{center}
\caption{Additional fields required to UV-complete the flavour sector of the deconstructed hypercharge model, and the mass or mixing observables they are responsible for generating.
}
\label{tab:fieldcontent2}
\end{table}

When discussing aspects of the UV completion, we have in mind a setup where NP states reside at two scales, separated by a mild hierarchy corresponding to a loop factor.
The first NP threshold is $f \approx$ TeV, where the $Z'$ resides, while at a higher scale $\Lambda_{\rm UV} \leq 4\pi f \approx 12$ TeV we assume there are new dynamics that 
\begin{enumerate}
    \item[(a)] Gives rise to the effective Yukawa operators for the light fermions (namely $Q$ and $H_{12}$), and
    \item[(b)] Solves the large hierarchy problem.
\end{enumerate}
Therefore, the scalar sector at the scale $f$ is technically natural, since it would receive radiative corrections from the high scale of size~\cite{Allwicher:2020esa}
\begin{equation}
\delta f^2  \approx \frac{\Lambda_{\rm UV}^2}{16\pi^2} \leq f^2.
\end{equation}
In this section, we will actually be led to a scenario whereby $\mathcal{G}_{\rm DH}$ is spontaneously broken {\em both} at $\Lambda_{\rm UV}$, via $\langle \Phi_q \rangle$, and at $f$ via $\langle \Phi_H \rangle$. Since the contribution of $\Phi_q$ to the $Z'$ mass is screened by the LH quark hypercharge $X_q$ (see Eq.~\ref{eq:Zp-mass}), both VEVs give comparable TeV scale contributions to the $Z'$ mass.

\subsection{Light--heavy mixing via a vector-like quark} \label{sec:light-heavy-mixing}

In order to generate the operators in~\cref{eq:LHd5}, which are responsible for $V_{cb}$ and $V_{ub}$, we add a vector-like quark with the same quantum numbers as $q_L^3$, namely $Q_{L,R} \sim ({\bf 3, 2}, 1/6,0)$. 
The relevant part of the renormalisable Lagrangian is
\begin{align}
\mathcal{L}_{\rm VLF} = & -m_Q \bar{Q}_L Q_{R}-\lambda_{q}^{i} \bar q_L^i \Phi_q Q_R - y_{+} \bar{Q}_L \tilde{H_3} t_R - y_{-} \bar{Q}_L H_3 b_R \,,
\end{align}
where we have chosen the basis where the mass mixing with $q_L^3$ has been rotated away and $m_Q$ is real. The other parameters are generally complex. The 2-vector of couplings $\lambda_q^i$ breaks the $U(2)_q$ flavour symmetry, and should therefore be of $O(V_{cb})$ if one wants to preserve the approximate $U(2)_q$ symmetry of the model. 

Integrating out $Q$ yields
\begin{equation}
\mathcal{L}_{d=5} \supset \frac{\lambda_q^i}{m_Q} \left(  y_{+} \bar q_L^i \Phi_q \tilde{H_3} t_R + y_{-} \bar q_L^i \Phi_q H_3 b_R \right) \,,
\label{eq:Ld5Vcb}
\end{equation}
matching onto the effective Lagrangian~\cref{eq:LHd5}. 
If the parameters $y_+$ and $y_-$ are of similar size, then the down sector dominates the CKM mixing due to an enhancement by the ratio $y_t/y_b$. The matching condition is then%
\begin{equation}
y_- \lambda_q^i \vq/m_Q = y_b V_{ib}.
\end{equation}
In fact, if we take $m_Q \sim 4\pi f$ in line with our general assumptions above, we will see later (\S~\ref{sec:up-alignment}) that a combination of flavour constraints and naturalness arguments suggests we consider this pure `up-aligned' scenario for 2-3 quark mixing.
Therefore, we will consider the scenario where the CKM comes dominantly from the down-sector whenever it is phenomelogically relevant.

We remark that an alternative option for generating the heavy-light mixing would be to add VLQs which have the same quantum numbers as $t_R$ and/or $b_R$ after DH symmetry breaking. However, with that choice, one also generates contributions to the light fermion Yukawas via couplings to $\Phi_H$. This would generate the light Yukawas via the product of two linear spurions $C^{ij}_{u,d} \sim \lambda_L^i \lambda_{u,d}^j$ \`a la partial compositeness, where $\lambda_{L,u,d}^i$ break the global $U(2)_{q}$ and $U(2)_{u,d}$ symmetries. This non-minimal breaking of the $U(2)^5$ symmetry via the non-SM-like spurions $\lambda_{u,d}$ leads in general to the generation of dangerous FCNCs via scalar operators and dipoles that push the NP to higher scales~\cite{Panico:2016ull}, spoiling the degree of naturalness we will ultimately achieve with our setup. For this reason, we stick with the $Q_{L,R}$ VLQ option, which breaks only $U(2)_{q}$.

Note also that since the leptonic mixing matrix may come entirely from the neutrino sector, there is no need to include the corresponding vector-like lepton (VLL); thus, in our UV completion, one does not in fact generate the light-heavy charged lepton mixing expected in Eq.~(\ref{eq:mu-tau}).\footnote{If one wishes to eventually embed $\mathcal{G}_{\rm DH}$ in a semi-simple gauge group by, say, unifying quarks and leptons via $SU(4)$ colour, then the VLL comes along with the VLQ, and one expects order-$V_{cb}$ sized mixing of left-handed muons and taus. The main phenomenological consequence of this is a tree-level contribution to the lepton-flavour-violating (LFV) $\tau \to 3 \mu$ decay. The current bound~\cite{ParticleDataGroup:2022pth} would even then give only a weak constraint on our model; we find $M_{Z'} \gtrsim 0.45\, \text{TeV}/\cos\theta$.}

\subsection{Light--heavy flavour violation constraints} \label{sec:light-heavy-FV}

Recall that our envisaged UV physics allows for a complicated NP sector at the higher threshold scale $4\pi f$, such as strong dynamics or SUSY, to protect the Higgs and the scale $f$ from UV physics  $> 4\pi f$. Even without specifying this sector explicitly, one can nonetheless use a spurion analysis to estimate the size of flavour-violating effects it can induce, given the source of flavour violation we have just introduced to explain $V_{cb,ub}$. 

As already mentioned, the 2-vector of couplings $\lambda_q^i$ is a doublet of $U(2)_q$ flavour symmetry, and so $\lambda_q^i = V_q^i$ acts as our flavour-breaking spurion.
With this, one can construct the following relevant invariants
\begin{equation}
\bar q_L^i V_q^i\gamma_\mu q_L^3\,, \hspace{12.5mm} \bar q_L^i V_q^i \sigma_{\mu\nu} b_R H\,.
\end{equation}
As we will see, both operators give relevant constraints. 

\subsubsection*{$\bm{B_{s(d)}}$ meson mixing}

We start with vector operators of the first type that mediate purely left-handed $B_{s,d}-\bar B_{s,d}$ mixing via the effective lagrangian
\begin{align}
\mathcal{L}_{d=6} &\supset \frac{c_B}{2\Lambda_B^2} (\bar q_L^i V_q^i \gamma_\mu q_L^3)^2\, \rightarrow C_{B_i}^1 (\bar d_L^i \gamma_\mu b_L)^2 \,, \hspace{17mm} C_{B_i}^1 \approx \frac{c_B}{2\Lambda_B^2} (V_{q}^{i})^2 \,.
\end{align}
The bounds we need to respect are $|C_{B_s}^1| \lesssim 10^{-5}~{\rm TeV}^{-2}$ and $|C_{B_d}^1| \lesssim 10^{-6}~{\rm TeV}^{-2}$ at 95\% CL~\cite{Gherardi:2020qhc,UTfit:2007eik,UTfit:latest}. Assuming $c_B \sim 1$, as one would expect in a strongly-coupled UV theory (for example, one in which the Higgs is composite\footnote{To be clear, when we refer to `strongly-coupled' and `weakly-coupled' UV completions, we are referring to models with a dynamical solution to the hierarchy problem in which FCNCs arise at tree-level or are loop-suppressed, respectively. We have in mind a composite Higgs model as indicative of the former scenario, and some weakly-coupled supersymmetric theory as an example of the latter, but this is not a sharp correspondence; for example, a SUSY model could feature tree-level flavour violation if not $R$-parity symmetric, in which case it would fall into our `strongly-coupled' class.}), and $\Lambda_B \sim 4\pi (\text{1 TeV}) \sim 12$ TeV, we require 
\begin{equation}
    \lambda_q^i = V_q^i \lesssim |V_{ti}| \quad \text{(strongly-coupled UV)} \, .
\end{equation}
On the other hand, for a weakly-coupled UV completion (for example, one with SUSY), it is reasonable for meson mixing amplitudes to be loop suppressed. Taking  $c_B \sim (16\pi^2)^{-1}$ and  $\Lambda_B \sim 12$ TeV, we find only that 
\begin{equation}
    \lambda_q^i = V_q^i \lesssim 4\pi |V_{ti}| \quad \text{(weakly-coupled UV)} \, ,
    \label{eq:VqBoundWC}
\end{equation}
which would allow for a much larger breaking of the $U(2)_q$ symmetry.

\subsubsection*{$\bm{B \to X_s \gamma}$ decays}

We now turn to the tensor structure $\bar q_L^i V_q^i \sigma_{\mu\nu} b_R H$, which when contracted with the SM gauge field strengths mediates $B\rightarrow X_s \gamma$ decays. Considering first the contributions from the lightest layer of BSM ingredients that we have explicitly specified in the model (Tables~\ref{tab:fieldcontent} and~\ref{tab:fieldcontent2}), such dipole operators are radiatively generated via a diagram with the $Z^\prime$ and VLQ in the loop. This contribution is proportional to $y_- V_q^2/(4\pi)^2$ -- doubly suppressed by both a loop factor and by small couplings, making it easily compatible with the experimental bounds. To go further, we should also address whether the unspecified layer of new physics at the high scale $4\pi f$ might give larger contributions to these observables, which we next discuss for generic classes of UV completion.

In general, there are three relevant structures
\begin{equation}
\mathcal{L}_{\rm SMEFT} \supset C_{dG} \,\mathcal{O}_{dG} + C_{dW} \,\mathcal{O}_{dW}+ C_{dB} \,  \mathcal{O}_{dB} \,,
\end{equation}
where $\mathcal{O}_{dG}$, $\mathcal{O}_{dW}$, and $\mathcal{O}_{dB}$ are defined as
\begin{equation}
{\cal O}_{dG}= (\bar q_L^2 \sigma^{\mu\nu} T^a b_{R} ) H G_{\mu\nu}^a \,, \quad {\cal O}_{dW}= (\bar q_L^2 \sigma^{\mu\nu} \tau^I b_{R} ) H W_{\mu\nu}^I \,,\quad {\cal O}_{dB}= (\bar q_L^2 \sigma^{\mu\nu}  b_{R} ) H B_{\mu\nu} \,.
\end{equation}
Because these operators all mix into the photon dipole under RGE, a precise analysis requires computing these Wilson coefficients in a particular model. To obtain a rough estimate, we assume all operators are generated at the high scale $\Lambda_B \sim 12$ TeV with the same Wilson coefficients (up to gauge couplings, which must appear differently in each case). Specifically, we take
\begin{equation}
C_{dG} = g_s \frac{c_B}{\Lambda_B^2} V_q^2\,, \hspace{10mm} C_{dW} = g_L \frac{ c_B}{\Lambda_B^2} V_q^2\,, \hspace{10mm} C_{dB} = g_Y \frac{c_B}{\Lambda_B^2} V_q^2\,,
\end{equation}
and follow the analysis in Ref.~\cite{Lizana:2023kei} for the RGE and computation of the corresponding photon dipole at the EW scale. Using the theory expressions given in~\cite{Misiak:2020vlo} for $B\rightarrow X_s \gamma$, we find a bound of $\Lambda_{B} \gtrsim \sqrt{c_B V_q^2} \times 150$ TeV at 95\% CL. If we instead demand $\Lambda_B \sim 4\pi (\text{1 TeV}) \sim 12$ TeV, this translates into following bounds on $V_q^2$
\begin{align}
    \lambda_q^2 = V_q^2 &\lesssim 0.007\, , \quad &&\text{(strongly-coupled UV)} \, ,\\
    \lambda_q^2 =V_q^2 &\lesssim 1\, , \quad &&\text{(weakly-coupled UV)} \,.
\end{align}
The takeaway from this analysis is that as long as these dipoles are loop-generated, $B\rightarrow X_s \gamma$ does not provide any significant constraint on the model.

However, in the case of a strongly-coupled UV (which could generate $c_B = 1$), $V_q^2$ should be smaller than $V_{ts}$ to be fully safe in the absence of additional suppression. That said, it is not unreasonable to expect the coupling between $b_R$ and the Higgs to be suppressed even in strongly-coupled UV completions. For example, we could take our cue from the partial compositeness paradigm, wherein the smallness of the bottom Yukawa is explained due to a (mostly) elementary $b_R$ field, which results in a small Yukawa coupling to the Higgs that lives in the composite sector. Such scenarios also lead to $y_b$-sized suppression in dipoles involving $b_R$ (which have the same structure as the Yukawas), bringing $B\rightarrow X_s \gamma$ well under control. Still, we note that $B\rightarrow X_s \gamma$ provides a condition on strongly-coupled UV completions, namely that there should be an additional suppression factor of around 6 to allow for $V_q^2 \sim V_{ts}$.

\subsection{Light Yukawas via a heavy Higgs}
\label{sec:lightYuks}

\begin{figure}
    \centering
    \hspace{-0.3cm}\includegraphics[width=0.85\columnwidth]{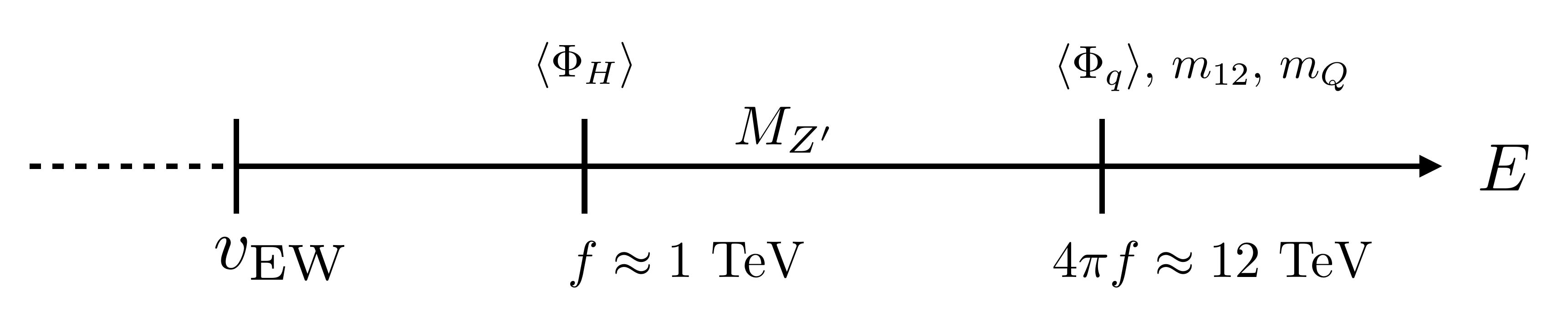}
    \caption{Diagram showing the distribution of scales in our setup. In particular $\vh$ lies near the scale $f \approx 1$ TeV, while $\vq$, $m_{12}$, and $m_Q$ lie at the high scale $4\pi f$. The $Z'$ mass comes from the combination $M_{Z'} \propto \sqrt{\vh^2 +\vq^2/9}$, and therefore lies somewhere in between $f$ and $4\pi f$.
    \label{fig:scaleDiagram}}
\end{figure}

To generate the light fermion Yukawa couplings in \cref{eq:LHd5light}, we extend the theory by a second $SU(2)_L$-doublet complex scalar field, charged under $U(1)_{12}$ rather than $U(1)_3$ (see Table~\ref{tab:fieldcontent2}). This allows renormalisable Yukawa couplings to be written down for the light fermions, but involving the $H_{12}$ state rather than $H_3$. Together, we have 
\begin{align} \label{eq:ren-yuk}
\mathcal{L}_{\rm Yuk} = &-y^3_u \bar q_L^3 \tilde{H_3} t_R -y^3_d \bar q_L^3 H_3 b_R  -y^3_e \bar \ell_L^3 H_3 \tau_R  \nonumber \\
 &-\bar q_L \hat Y_u\tilde{H}_{12} u_R - \bar q_L O_{12} \hat Y_d H_{12} d_R  - \bar \ell_L \hat Y_e H_{12} e_R \,,
\end{align}
where without loss of generality we have chosen a physical basis where $y^3_{u,d,e}$ are real numbers, $\hat Y_{u,d,e} = {\rm diag}(y^{1}_{u,d,e},\, y^{2}_{u,d,e})$ are $2\times 2$ real diagonal matrices, and $O_{12}$ is an orthogonal $2\times2$ matrix parameterized by an angle $\theta_{12}$. The fact that the couplings of the heavy Higgs $\tilde H_{12}$ to light quarks and leptons can be taken to be purely real is an important phenomenological feature of the model; it means that these couplings do not introduce new physical sources of $CP$-violation in the light fermion sector -- as we explore further below.\footnote{In the model as written, any dipoles involving light leptons must be proportional to $\hat Y_e$, as it is the only source of $U(2)_\ell \times U(2)_e$ symmetry breaking -- as emphasized above, we do not add a VLL (along with the VLQ), that would provide an additional spurion transforming in the doublet of $U(2)_l$. Since $\hat Y_e$ can always be taken real and diagonal, the model predicts no charged lepton flavor violation or electric dipole moments involving the light leptons.}

The two Higgs fields $H_{12}$ and $H_3$ mix via interactions in the scalar potential. The key interaction, as we see in more detail in \S~\ref{sec:scalar-potential}, is a cubic interaction $V \supset -f H_{12}^\dagger \Phi_H H_3$. The physical mass eigenstates $H_h$ and $H_l$ are then linear combinations of $H_{12}$ and $H_3$, rotated by a small mixing angle $\tan\delta_H \approx f \langle \Phi_H \rangle / m_{12}^2 \ll 1$, where $m_{12}$ is the mass of $H_{12}$. The component $H_l$, which is mostly $H_3$, is identified with the physical Higgs boson whose mass squared emerges as the result of a percent level tuning (see \S~\ref{sec:scalar-potential}). Indeed, this can also be seen by integrating out the heavy Higgs $H_{12}$. Keeping operators up to dimension-5, we obtain
\begin{equation}
\mathcal{L}_{d\leq 5} \supset \frac{f}{m_{12}^2} \left(f |\Phi_H|^2 |H_3|^2 - \bar q_L \hat Y_u \Phi_H^* \tilde{H}_{3} u_R - \bar q_L O_{12} \hat Y_d \Phi_H H_{3} d_R  - \bar \ell_L \hat Y_e \Phi_H H_{3} e_R  \right) \,,
\label{eq:intOutH12}
\end{equation}
where $H_3 = H_l$ in this limit. Therefore, we see that the renormalisable interactions written in the second line of~\cref{eq:ren-yuk} are effectively dimension-5 suppressed interactions with the physical Higgs $H_l$ (where the relevant heavy scale is here $\Lambda_H = m_{12}^2/f \approx 16\pi^2 f$).\footnote{The first term in \cref{eq:intOutH12} corrects the mass of $H_l$ after DH symmetry breaking and will be discussed in detail along with all other Higgs mass corrections in \S~\ref{sec:scalar-potential}.} From \cref{eq:ren-yuk,eq:intOutH12}, we see that
the effective charm Yukawa satisfies
\begin{equation}
\frac{y_c}{y_t} = \frac{y_u^2}{y_u^3} \tan\delta_H \approx  \frac{y_u^2}{y_u^3}  \frac{f \langle \Phi_H \rangle}{m_{12}^2} \,,
\end{equation}
and at $1$ TeV we have~\cite{Xing:2007fb}:
\begin{equation}
\left(\frac{y_c}{y_t}\right)_{\mu = 1\,{\rm TeV}} = (3.5 \pm 0.5) \times 10^{-3} \,,
\end{equation}
so we see that the hierarchy between $y_t$ and $y_c$ can be explained for fundamental Yukawa couplings $y_u^2 \approx y_u^3 \approx 1$ if $\langle \Phi_H \rangle \sim f$ and $m_{12} \sim 4\pi f$. This indirectly leads to a constraint on the $Z'$ boson mass, and both symmetry breaking scales $\langle \Phi_{q,H} \rangle$, as follows.

Since we want $f$ around 1 TeV to minimize the little hierarchy problem, we see that the relation $\langle \Phi_H \rangle \approx f$, together with the $Z'$ mass formula in \cref{eq:Zp-mass} would lead to a $Z'$ that is too light unless it receives a significant contribution to its mass from $\vq$. Therefore, we are pointed to a r\'egime where 
\begin{equation} \label{eq:vev_scales}
    \vh \approx f, \qquad  \vq \approx m_{12}\approx 4\pi f,
\end{equation} 
such that we explain $y_c/y_t \approx (16\pi^2)^{-1}$ while also having a $Z'$ with a (phenomenologically viable) mass of a few TeV. In practice, given the freedom to adjust $m_{12}$ by an $O(1)$ factor, we will consider $ \vh /f  \in [1,4]$ which we consider to provide a satisfactory explanation of $y_c/y_t$ without placing a strong requirement on the size of $|\Phi_H|^2 |H_3|^2$ cross-quartics. The distribution of the various scales in our model is summarised graphically in \cref{fig:scaleDiagram}.

\subsection{Light family flavour violation constraints} \label{sec:light-flavour-V}

Given our spurion content, there are two important structures that control light family flavour violation. The first can be built from the $2\times 2$ light family Yukawa couplings $Y_u Y_u^{\dagger}$, where $Y_u = O_{12}^T \hat Y_u$ in the light-quark down basis such that $Y_u Y_u^{\dagger}$ is purely real. The typical expectation is that flavor violation from $Y_u Y_u^{\dagger}$ is suppressed by the charm Yukawa. However, it is important to emphasize that this is not the case in our setup, where $Y_u Y_u^{\dagger}$ is instead proportional to $y_u^2$, which we would like to be an $O(1)$ parameter in order to explain $m_c/m_t$ via a hierarchy of scales. The relevant second structure is $V_q V_q^\dagger$, which in general is complex and indeed must be so in order to generate the CKM phase. Thus, even though the expected size of $V_q V_q^\dagger$ is parameterically smaller than $Y_u Y_u^{\dagger}$, its phases lead to competitive bounds from CP-violating observables such as $\epsilon_K$.

\subsubsection*{Kaon mixing (real part)}

Let us first analyze $Y_u Y_u^{\dagger}$, where we can construct MFV-like operators in the EFT with two relevant FCNC structures
\begin{equation}
\bar q_L^i [Y_u Y_u^{\dagger}]_{ij}\gamma_\mu q_L^j\,, \hspace{10mm} \bar q_L^i [Y_u Y_u^{\dagger} \hat Y_d]_{ij} \sigma_{\mu\nu} d_R^j \,.
\end{equation}
The tensor operator generates flavour-violating dipoles mediating $s \rightarrow d \gamma$ which are less constrained than vector operators of the first type that mediate purely real and left-handed $K-\bar K$ mixing
\begin{align}
\mathcal{L}_{d=6} &\supset \frac{c_K}{2\Lambda_K^2} (\bar q_L^1 [Y_u Y_u^{\dagger}]_{12}\gamma_\mu q_L^2)^2\, \hspace{2mm}\longrightarrow \hspace{2mm} C_K^1 (\bar d_L \gamma_\mu s_L)^2\,, \hspace{7.5mm} C_K^1 \approx \frac{c_K}{2\Lambda_K^2} c_{12}^2 s_{12}^2 (y_u^2)^4 \,,
\end{align}
where \emph{e.g.} $s_{12} = \sin \theta_{12}$. The bound we need to respect is $|{\rm Re}(C_K^1)| \lesssim 10^{-6}~{\rm TeV}^{-2}$ at 95\% CL~\cite{Gherardi:2020qhc,UTfit:2007eik,UTfit:latest}. 
Again, one would interpret this bound a little differently depending on the class of UV completion; strong dynamics suggests $c_K \sim 1$, while weak dynamics would be consistent with a 1-loop suppresion, $c_K \sim (16\pi^2)^{-1}$. As before, taking $\Lambda_K \sim 4\pi(\text{1 TeV}) \sim 12$ TeV we obtain
\begin{align}
    y_u^2 &\lesssim 0.3\, , \quad &&\text{(strongly-coupled UV)} \, ,\\
    y_u^2 &\lesssim 1\, , \quad &&\text{(weakly-coupled UV)} \, ,
\end{align}
and so order-1 values for $y_u^2$ can be well-accommodated.

\subsubsection*{CP violation in kaon mixing}

Moving now to the $V_q V_q^\dagger$ structure, again we can build operators such as 
\begin{align}
\mathcal{L}_{d=6} &\supset \frac{c_K}{2\Lambda_K^2} (\bar q_L^1 [V_q V_q^{\dagger}]_{12}\gamma_\mu q_L^2)^2\, \hspace{2mm}\longrightarrow \hspace{2mm} C_K^1 (\bar d_L \gamma_\mu s_L)^2\,, \hspace{7.5mm} C_K^1 \approx \frac{c_K}{2\Lambda_K^2}([V_q V_q^{\dagger}]_{12})^2 \,,
\end{align}
that give a complex NP contribution to kaon mixing. The constraining observable in this case is therefore $\epsilon_K$, which gives a bound of $|{\rm Im}(C_K^1)| \lesssim 10^{-9}~{\rm TeV}^{-2}$ at 95\% CL~\cite{Gherardi:2020qhc,UTfit:2007eik,UTfit:latest}. 
As before, looking at $c_K = 1$ and $(16\pi^2)^{-1}$ and taking $\Lambda_K \sim 4\pi(\text{1 TeV}) \sim 12$ TeV we obtain
\begin{align}
    |{\rm Im}([V_q V_q^{\dagger}]_{12})| &\lesssim 6\times 10^{-4} \, , \quad &&\text{(strongly-coupled UV)} \, ,\\
    |{\rm Im}([V_q V_q^{\dagger}]_{12})| &\lesssim 7\times 10^{-3}\, , \quad &&\text{(weakly-coupled UV)} \, ,
\end{align}
where for reference we have $|{\rm Im}(V_{td}^{*} V_{ts})| \approx 10^{-4}$. We therefore require ${\rm Im}([V_q V_q^{\dagger}]_{12})$ to be ``CKM-like" within a factor of 6 in the case of a strongly-coupled UV (where $c_K$ could be unsuppressed), while in the weakly coupled case we can gain an extra factor of $4\pi$ in the size of the product (or perhaps $\sqrt{4\pi}$ in $V_q^i$ themselves). Note that this is more restrictive than the bound in~\cref{eq:VqBoundWC}, due to the quartic scaling of kaon mixing with $V_q^i$.

\subsubsection*{$\bm{K^+ \to \pi^+ \nu \bar{\nu}}$ decays}

It is not only 4-quark operators that give relevant flavour bounds here.
One can also build the semi-leptonic operator
\begin{align}
\mathcal{L}_{d=6} &\supset \frac{c_{\alpha}}{\Lambda_{K\pi}^2} (\bar q_L^i [Y_u Y_u^{\dagger}]_{ij}\gamma_\mu q_L^j)(\bar \ell_L^\alpha \gamma^\mu \ell_L^\alpha) 
\hspace{4mm}\longrightarrow \hspace{4mm} C^{\nu}_\alpha (\bar d_L \gamma_\mu s_L)(\bar \nu_L^\alpha \gamma^\mu \nu_L^\alpha)\,,
\end{align}
for any lepton flavour $\alpha = 1,2,3$. This would mediate $K^+\rightarrow \pi^+ \nu\bar\nu$ transitions with only one power of the Cabibbo angle for suppression.\footnote{Contributions from $V_q V_q^\dagger$ are also possible, but are sub-leading in size compared to $Y_u Y_u^{\dagger}$ since $K^+\rightarrow \pi^+ \nu\bar\nu$ is sensitive to the real part of these objects.} Namely, we have
\begin{equation}
C^{\nu}_\alpha \approx -\frac{c_\alpha}{\Lambda_{K\pi}^2} c_{12} s_{12} (y_u^2)^2 \,.
\end{equation}
Assuming a lepton universal contribution ($C^{\nu}_e = C^{\nu}_\mu = C^{\nu}_\tau$), the bound we need to pass is $|C^{\nu}_\alpha| \lesssim 10^{-4}~{\rm TeV}^{-2}$~\cite{Crosas:2022quq}. Taking $\Lambda_{K\pi}^2 \sim 12$ TeV, we find the same numerical bounds as from $K-\bar K$ mixing, namely $y_u^2 \lesssim 0.3 (1.0)$ for strong (weak) UV, but now without the ambiguity from long-distance effects present in $\Delta M_K$.

\section{Naturalness and Constraints} \label{sec:naturalness}

So far, we have set out an economical model of flavour in which a deconstructed hypercharge gauge symmetry delivers $U(2)^5$ accidental symmetries, which are minimally broken by a VLQ $Q$ and a second Higgs doublet $H_{12}$ in order to generate the light Yukawa couplings. The model entails a small number of BSM states that couple to the SM Higgs with various strengths, leading to radiative corrections to the Higgs mass. In this section, we estimate the extent to which our setup is natural, finding a reasonable parameter space in which the tuning in $m_h^2$ is no worse than the unavoidable `little hierarchy' corresponding to the fact that $m_h \sim 100$ GeV rather than its natural scale of $m_h \sim f \sim 1$ TeV.

\subsection{Scalar potential and `tree-level' tuning} \label{sec:scalar-potential}
The scalar sector of our model consists of the electroweakly-charged Higgs doublets $H_3$ and $H_{12}$, together with two $SU(2)_L$ singlet scalars $\Phi_H$ and $\Phi_q$ whose VEVs together contribute to the breaking $U(1)_{12} \times U(1)_3 \to U(1)_Y$. The most general renormalisable scalar potential involving these four fields is
\begin{align}
V \supset & -m_{3}^2 |H_3|^2 +\lambda
_{3}|H_{3}|^4 - f H_{12}^{\dagger}\Phi_{H} H_{3} + m_{12}^2 |H_{12}|^2 +\lambda_{12} |H_{12}|^4 +\lambda_{123} |H_{12}|^2 |H_3|^2 \nonumber \\
&+ \kappa_{H} \big[ |\Phi_H|^2 - \langle \Phi_H \rangle^2 \big]^2
 + \kappa_{q} \big[ |\Phi_q|^2 - \langle \Phi_q \rangle^2 \big]^2 +\kappa_{Hq} |\Phi_H|^2 |\Phi_q|^2 - \kappa_{Hqqq}\, \Phi_H^{*}\Phi_{q}^3 \nonumber \\
& + \left(\kappa_{3H} |\Phi_H|^2 + \kappa_{3q} |\Phi_q|^2 \right) |H_3|^2  + \left(\xi_{12H} |\Phi_H|^2 + \xi_{12q} |\Phi_q|^2 \right) |H_{12}|^2  + {\rm h.c.}  \,, \label{eq:V}
\end{align}
where one can always re-phase $\Phi_{H,q}$ to make the couplings $f$ and $\kappa_{Hqqq}$ real. Generically in a scenario where two scalar fields together break a single gauged $U(1)$, there is na\"ively a massless Goldstone mode $\varphi_A$ that goes uneaten.\footnote{Specifically, defining $\Phi_a = \rho_a e^{i \varphi_a / \langle \Phi_a \rangle}$, the Goldstone modes are $\varphi_{Z'} = \varphi_q \cos\beta  + \varphi_H \sin\beta$ and $\varphi_{A} = \varphi_H \cos\beta  - \varphi_q \sin\beta$,
where the state $\varphi_{Z'}$ is the Goldstone eaten by the $Z'$.} However, DH gauge invariance allows for the quartic coupling $\kappa_{Hqqq}$ that breaks the global symmetry associated to the massless mode. This explicit breaking leads to a mass for $\varphi_A$ of
\begin{equation}
M_{A}^2 = 2 \kappa_{Hqqq}  \frac{\vh^2}{\cos^2\beta} \frac{\vq^3}{\vh^3} \,,
\end{equation}
where $\tan\beta = 3 \vh /\vq $. Because the quartic coupling $\kappa_{Hqqq}$ is a symmetry breaking parameter, it is technically natural for it to be small; we assume it is such that $\kappa_{Hqqq} \leq (4\pi)^{-3}$, in order to not destabilize the VEV hierarchy $\vh \approx f
\approx 1$ TeV $\ll \vq \approx 4 \pi f$, following~\cref{eq:vev_scales}. This gives $M_A \approx \sqrt{2}\vh$, so this value for $\kappa_{Hqqq}$ does not result in a very light state.
We also assume a hierarchy in the mass scales $m_{12} \approx 4\pi f \gg m_{3} \approx O(f)$.  Finally, concerning the other quartics in~\cref{eq:V}, all can be $O(1)$ except for $\kappa_{3q}$ and $\kappa_{Hq}$, which we assume to be generated only radiatively in order not to give large corrections to the $H_3$ mass (in the case of $\kappa_{3q}$) or to destabilize $\langle \Phi_q \rangle / \langle \Phi_H \rangle \approx 4\pi$ (in the case of $\kappa_{Hq}$). One possible way to justify these hierarchies could be if $\Phi_H$ and $H_3$ are realised as pseudo Nambu-Goldstone bosons, which could point towards a strongly-coupled theory at the high scale $4\pi f$. Here, we take an agnostic approach where we simply note that these choices are technically natural in the sense that they are radiatively stable.
 
 After SSB of $\mathcal{G}_{\rm DH}$ to the SM, a mass mixing is induced between $H_{3}$ and $H_{12}$. Diagonalizing the system, they are related to the mass eigenstates $H_{h}$ and $H_l$ by
\begin{align}
H_{3} &= H_l \cos \delta_H  - H_{h} \sin \delta_H \,, \\
H_{12} &= H_{h} \cos \delta_H  + H_l \sin \delta_H \,,
\end{align}
where the mixing angle is $\tan \delta_H \approx f\langle \Phi_H \rangle / m_{12}^2 $ and $H_l$ is identified as the SM Higgs doublet. For small $\delta_H$, the tree-level mass eigenvalues are
\begin{align}
M_{H_{l}}^2 &\approx \kappa_{3H} \langle \Phi_H \rangle^2 + \kappa_{3q} \langle \Phi_q \rangle^2 - \delta_H^2 m_{12}^2 -m_3^2  \,, \label{eq:treeHiggsMass} \\
M_{H_{h}}^2 &\approx \xi_{12H} \langle \Phi_H \rangle^2 + \xi_{12q} \langle \Phi_q \rangle^2 + (1+\delta_H^2) \, m_{12}^2   \,,
\end{align}
so we see that we need to fine tune $m_3^2$ such that $M_{H_{l}}^2 \approx -(100\, {\rm GeV})^2$ instead of its natural value of $O(f^2)$. This means we can also allow for quantum corrections to $M_{H_{l}}^2$ that do not exceed $f^2$. This is the irreducible little hierarchy problem corresponding to an order $1\%$ fine tuning. Once this is done, only $H_{l}$ has a tachyonic mass and acquires a non-zero VEV. 

\subsection{Radiative corrections} \label{sec:natural-1loop}

Our DH model of flavour features an extra $Z'$ gauge boson, a VLQ, and heavy scalar fields $H_{12}, \Phi_H, \Phi_q$. All these states are heavier than the SM Higgs and therefore contribute loop corrections to $M_{H_l}^2$ which are quadratically divergent, with the naive cutoff expected to be given by the particle mass squared. Along these lines, we follow the `finite naturalness' paradigm~\cite{Farina:2013mla}, by which we ignore incalculable divergences and seek to control the finite and log-divergent $\delta M_{H_l}^2$ contributions $\propto M_{\rm NP}^2$. Specifically, we require $\delta M_{H_l}^2 < f^2$ in order to not to worsen the tree-level little hierarchy problem defined in the previous section. As we have already emphasized, the idea is that unspecified dynamics solving the large hierarchy problem at the high scale $ 4\pi f$ will shield $M_{H_l}^2$ from NP in the deep UV, resulting only in finite contributions $\delta M_{H_l}^2 \lesssim f^2$, i.e. of the same size as the ones we will compute here.

We now compute 1-loop radiative corrections to the Higgs mass $M_{H_l}$ from all heavy states in our model. We calculate all loops in the SM symmetric but $\mathcal{G}_{\rm DH}$ broken phase, using the $\overline{\text{MS}}$ renormalization scheme. This means all fields should first be rotated to their mass eigenstates after $\mathcal{G}_{\rm DH}$ symmetry breaking before computing loops. All results here were computed using \texttt{PackageX}~\cite{Patel:2015tea} cross-checked by the \texttt{Machete}~\cite{Fuentes-Martin:2022jrf} 1-loop matching software.

\subsubsection*{Gauge boson Higgs mass corrections}

The 1-loop Higgs mass correction from the $Z'$ reads
    \begin{equation}
    \delta M_{H_l}^2(Z') = \frac{g_{Y}^2 X_H^2}{16\pi^2} \frac{M_{Z'}^2}{\tan^2\theta} \left[1+3\log\frac{\mu^2}{M_{Z'}^2}\right]\,,
    \end{equation}
    where recall $X_H$ is the SM hypercharge of the Higgs. One should take $\mu^2 \sim \Lambda_{\rm UV}^2 = (4\pi f)^2$ inside the argument of the logarithm in order to capture the leading-log renormalisation group evolution (RGE) of $M_{H_l}^2$ from the cutoff of the theory at $4\pi f$ down to $M_{Z'}$ (where the $Z'$ is no longer a light degree of freedom and can be integrated out). As anticipated in the introduction, one benefits from a suppression by the hypercharge prefactor $g_Y X_H \sim 0.15$, in addition to the 1-loop suppression, which allows the deconstructed hypercharge gauge boson to be naturally heavier than corresponding gauge bosons that would come from deconstructing $SU(2)_L$. The constraint on the DH model parameter space resulting from the $Z'$ radiative Higgs mass correction is shown in \S~\ref{sec:combResults}.

\subsubsection*{VLF Higgs mass corrections}
Any coloured VL fermion with a direct Higgs Yukawa coupling involving also one massless SM fermion gives a 1-loop Higgs mass correction with the following structure
\begin{equation}
\delta M_{H_l}^2({\rm VLF}) = -\frac{N_c}{8\pi^2} |y_{H\psi}|^2 M_{\psi}^2 \left[1+\log \frac{\mu^2}{M_\psi^2} \right]\,,
\label{eq:dmHVLF}
\end{equation}
where $y_{H\psi}$ is the coupling between the VLF and the SM Higgs in the VLF mass basis. Taking $M_{\psi} = 4\pi f = \mu$ and requiring $|\delta M_{H_l}^2({\rm VLF})| \leq f^2$, we find the following simple bound
\begin{equation} \label{eq:natural-VLF}
|y_{H\psi}| \leq \frac{1}{\sqrt{2N_c}} \approx 0.4 \,.
\end{equation}
This naturalness bound, combined with $\lambda_q^2 \lesssim V_{cb}$ from the spurion analysis (and taking $\vq = 4\pi f$), tells us that the CKM cannot come dominantly from the up-quark sector (See the discussion in \S~\ref{sec:light-heavy-mixing}).

\subsubsection*{Scalar Higgs mass corrections}
The most dangerous scalar Higgs mass corrections are quadratically divergent diagrams coming from cross-quartics with $H_l$ involving the heaviest fields, such as $|H_h|^2 |H_l|^2$ and $|H_l|^2 |\Phi_q|^2$.\footnote{There are also scalar cubics which generate only logarithmically divergent Higgs mass corrections proportional to either $f^2$, $\vq^2$, or $\vh^2$. These are never important compared to the tree-level Higgs mass corrections already induced by the same couplings, see~\cref{eq:treeHiggsMass} for more details.}   In practice, since $|H_l|^2 |\Phi_q|^2$ contributes to the $H_l$ mass already at tree-level via $\vq$ (see~\cref{eq:treeHiggsMass}), only $\lambda_{hl} |H_h|^2 |H_l|^2$ gives potentially important loop corrections. We find
\begin{equation}
\delta M_{H_l}^2(\lambda_{hl}) = -\frac{\lambda_{hl}}{8\pi^2}  M_{H_h}^2 \left[1+\log \frac{\mu^2}{M_{H_h}^2} \right]\,.
\end{equation}
Taking $\mu \approx M_{H_h}\approx 4\pi f$, there is no log-enhancement and we only need to require a weak bound on the coupling $\lambda_{hl} \lesssim 0.5$.

\subsection{Justification for up-alignment} \label{sec:up-alignment}

Finally, armed with these naturalness conditions, we can justify the assumption of `up-alignment' (whereby the mixing angles between light and third family quarks originates from the down-type Yukawa), that we anticipated in \S \ref{sec:light-heavy-mixing}.

Recall that the light--heavy mixing in the Yukawa sector comes in our model from dimension-5 operators in Eq. (\ref{eq:Ld5Vcb}), which are induced by integrating out the VLQ field $Q$.
Using (i) the assumption $m_Q \sim 4\pi f$; (ii) the naturalness upper bound $y_\pm \lesssim 0.4$ on the couplings of the VLFs (see \S~\ref{sec:natural-1loop}); (iii) that flavour constraints imply $\lambda_q^2 \lesssim |V_{cb}|$, at least, if the UV is strongly-coupled (see \S~\ref{sec:light-heavy-FV}); and (iv) that stability of the scalar potential means $\vq \lesssim 4\pi f$ (see \S~\ref{sec:scalar-potential}), we deduce that the contribution of the up-quark sector to the 2-3 CKM mixing is bounded as
\begin{equation}
\frac{y_{+}}{y_t} \frac{\lambda_q^2 \vq}{m_Q} \lesssim (0.4/y_t) V_{ts}\, , \nonumber
\end{equation}
meaning that one cannot explain the size of CKM mixing with pure down-alignment. It follows that light-heavy CKM mixing must receive a dominant contribution from the down-quark sector. Therefore, it is a convenient (and natural) approximation to assume pure up-alignment. This will also simplify our phenomenological analysis in \S~\ref{sec:pheno}, to which we now turn.

\section{Results and Phenomenology} \label{sec:pheno}

Naturalness pushes our model of flavour towards parameter space regions where the masses of new particles are near the TeV scale. As a result, there are relevant phenomenological consequences in high $p_T$ searches at the LHC, precision electroweak observables, and precision flavour observables, which we analyse in this Section.
\subsection{Tree-level SMEFT matching}
As emphasized throughout the paper, there is a well-defined (and natural) limit of the model in which the $Z'$ boson dominates the low-energy phenomenology, with the other states (the VLQ and heavy Higgs) being decoupled to the high scale $4\pi f$.\footnote{For a phenomenological overview of two Higgs doublet models with approximate $U(2)^5$ flavour symmetries, see \emph{e.g.}~\cite{Altmannshofer:2018bch,Botella:2016krk,Altmannshofer:2015esa,Blechman:2010cs,Altmannshofer:2012ar}.} The mass and couplings of the $Z'$ are given in Eqs. (\ref{eq:Zp-mass}) and (\ref{eq:Zp-couplings}). As justified above, we assume the `up-alignment' scenario for CKM mixing, {\em i.e.} that the up-quark Yukawa matrix is diagonal in the gauge interaction eigenbasis. It is then convenient to define the left-handed quark doublets as $q_{L}^i=(u_L^i, V_{ij} d_L^j)$, which means there is no flavour violation at the level of SMEFT operators.

The non-zero operators obtained by integrating out the $Z'$ at $\Lambda = M_{Z'}$ and matching onto the SMEFT at dimension-6 are as follows, in the Warsaw basis~\cite{Grzadkowski:2010es}. (As before, $X_\psi$ denotes the SM hypercharge quantum number of fermion species $\psi$, in a normalisation where $X_q=1/6$.)
First, the non-zero Wilson coefficients (WCs) for four-quark operators are
\begin{align}
    &\left(C_{qq}^{(1)}, C_{uu}, C_{dd}, C_{ud}^{(1)}, C_{qu}^{(1)}, C_{qd}^{(1)}\right)^{\alpha\alpha\beta\beta}  
    =-\left(X_q^2,X_u^2,X_d^2,2X_u X_d,2 X_q X_u,2 X_q X_d\right) \frac{g_{Z'}^{\alpha\alpha} g_{Z'}^{\beta\beta} }{2M_{Z'}^2}, 
\end{align}
where $\alpha,\, \beta \in \{1,2,3\}$ are flavour indices. 
All the four-lepton operator WCs are
\begin{align}
    \left(C_{ll}, C_{ee}, C_{le}\right)^{\alpha\alpha\beta\beta} =
    -\left(X_l^2,X_e^2,2 X_l X_e\right) \frac{g_{Z'}^{\alpha\alpha} g_{Z'}^{\beta\beta} }{2M_{Z'}^2}\, ,
\end{align}
while the non-zero semi-leptonic operators are
\begin{align} \label{eq:semi-leptonics}
    \left(C_{lq}^{(1)}, C_{eu}, C_{ed}, C_{lu}, C_{ld},C_{qe}\right)^{\alpha\alpha\beta\beta} 
    =-\left(X_l X_q,X_e X_u,X_e X_d,X_l X_u,X_l X_d,X_q X_e\right)\frac{g_{Z'}^{\alpha\alpha} g_{Z'}^{\beta\beta} }{M_{Z'}^2} \, .  
\end{align}
Next, we have the Higgs-bifermion operators:
\begin{align}
    \left(C_{H l}^{(1)}, C_{H e}, C_{H q}^{(1)}, C_{H u}, C_{H d}\right)^{\alpha\alpha}
    =-X_H\cot\theta \left(X_l,X_e,X_q,X_u,X_d\right)  \frac{g_{Z'}^{\alpha\alpha} }{M_{Z'}^2}\,.
    \label{eq:HiggsFermion}
\end{align}
Lastly, the purely bosonic operators are:
\begin{equation}
C_{H D} = 4 C_{H \Box} = -2 X_H^2 \cot^2 \theta \frac{g_Y^2}{M_{Z'}^2}\, .    
\label{eq:bosonic}
\end{equation}
 Note that the 4-fermion operators involving one light generation and one third generation fermion are independent of the gauge mixing angle $\theta$, as are the Higgs-bifermion operators that couple to {\em light} fermions (the Higgs is essentially a third family particle).

\subsection{Flavour-conserving constraints}
Here we analyse flavour-conserving constraints, which come dominantly from high-$p_T$ searches at ATLAS and CMS, as well as from electroweak precision tests, due to the direct NP couplings to the Higgs present in our DH model.

\subsubsection*{High-$p_T$ constraints}
Due to the $U(2)^5$ flavour symmetry of the DH gauge sector, LHC searches for the $Z'$ gauge boson must focus not only on di-electron and di-muon final states, i.e. $pp \rightarrow \ell\ell$ (with $\ell = e, \mu$) , but also di-tau final states $pp \rightarrow \tau\tau$. While light leptons generally give the strongest bounds, di-tau searches can be important for small values of the gauge mixing angle $\tan\theta$, where the $Z'$ couples mostly to third-family fermions and is therefore produced mainly through the process $bb \rightarrow Z' \rightarrow \tau\tau$ (perhaps also with an initial state $b$-jet from gluon splitting). 

In light of this, we use the \texttt{HighPT} package~\cite{Allwicher:2022mcg} to fully implement our model and compute both the $\sigma(pp \rightarrow ee+\mu\mu)$ total cross section, as well as $\sigma(pp \rightarrow \tau\tau)$. The \texttt{HighPT} package allows us to have analytic results for the cross section, up to a requirement of specifying numerical values of the $Z'$ mass and decay width in order to perform phase space integrals. Nevertheless, for a given value of $M_{Z'}$, we can obtain results for the cross section which are analytic in $\tan\theta$. Our strategy to obtain high-$p_T$ limits on the model is to compare our DH model prediction for a given cross section to the bounds given in~\cite{CMS:2021ctt} (for $ee+\mu\mu$) or~\cite{ATLAS:2017eiz} (for $\tau\tau$) to obtain the limit on $\tan\theta$ for each value of $M_{Z'}$. Note that while both searches have the same center of mass energy $\sqrt{s} = 13$ TeV, the ATLAS $\tau\tau$ search has only 36 $\text{fb}^{-1}$ compared to 140 $\text{fb}^{-1}$ for the CMS $\ell\ell$ search. We therefore perform a na\"ive rescaling of the bound on the $\tau\tau$ cross section by $\sqrt{36/140}$, to estimate the bound one would obtain with 140 fb$^{-1}$ of integrated luminosity.

\subsubsection*{Electroweak precision tests}
\label{sec:EWPO}
The Higgs-bifermion and fully bosonic operators generated after integrating out the $Z'$ at tree-level are constrained mainly by electroweak precision obervables (EWPOs). Since the Higgs-bifermion operators involving the light families (see~\cref{eq:HiggsFermion}) are independent of $\tan\theta$, decreasing the gauge mixing angle only results in larger contributions to EWPOs via third-family fermions and the fully bosonic operators in~\cref{eq:bosonic}. All these effects scale as $\cot^2\theta$, so avoiding high-$p_T$ bounds via new physics dominantly coupled to the third family inevitably leads to large corrections in EWPOs.  

On the fully bosonic side, the custodial symmetry breaking operator $C_{HD}\propto -X_H^2\cot^2\theta$ is of particular interest, since it has a fixed sign that leads to a large positive shift in the $W$-boson mass. Even before the recent CDF II measurement of $M_W$~\cite{CDF:2022hxs}, the EW fit could be improved by a non-zero (negative) value of $C_{HD}$, mainly due to a mild $2\sigma$ tension between the experimental determination of $M_W^{\rm old} = 80.379(12)$ GeV and the SM theory prediction~\cite{Breso-Pla:2021qoe}. A combination of $M_W$ measurements cannot now be done in a straightforward manner, due to the statistical incompatibility of CDF II with the others. Since work towards an official combination is still underway~\cite{LHC-TeVatronW-bosonmasscombinationworkinggroup:2022nay}, in the interim we apply the PDG approach where we penalize all measurements `democratically'. Specifically, we inflate all errors equally until the $\chi^2$ per degree of freedom is equal to one, yielding $M_W^{\rm new} = 80.410(15)$ GeV. To be conservative, we use $M_W^{\rm old}$ when obtaining exclusion limits from the EW fit, but we also show that this does not exclude the region preferred by $M_W^{\rm new}$. This is due to the fact that EW exclusion limit is dominantly determined by $Z\rightarrow  e_R\, e_R $ ($\tan\theta \approx 1$) and $Z\rightarrow \tau_R \, \tau_R $ ($\tan\theta < 1$) vertex corrections, due to right-handed charged leptons having the largest hypercharge. Indeed, as we will see, the EW fit prefers a non-decoupling behaviour for NP when using $M_W^{\rm new}$, since it calls for a large negative $C_{HD}$ contribution that is not excluded by the rest of the EW fit. We emphasize that, in this model, one cannot turn on $C_{HD}$ without turning on all the other relevant operators (such as $C_{He}$), because the ratios between the WCs are always ratios of hypercharges which are fixed. This results in a correlated set of NP effects that cannot be disentangled by tuning parameters. 

We take constraints from EWPO into account by performing the full EW fit using the same $(G_F, \alpha, m_Z)$ input scheme, observables, SM theory predictions, and likelihood as in~\cite{Breso-Pla:2021qoe}, which was previously validated and found to give accurate results in~\cite{Allwicher:2023aql}.  We also include leading-log running in $y_t$, $g_L$, and $g_Y$ of all operators generated by the $Z'$ at tree-level that mix via RGE into operators relevant for the EW fit. However, we find that the effect of including RGE is very small, which is due in large part to a cancellation of the dominant $y_t$ running for the most important operators; namely, we have that~\cite{Jenkins:2013wua}
    \begin{equation}
        \dot C_{He, Hl, Hd, HD} \propto y_t^2 (X_q - X_u + X_H)\,,
    \end{equation}
which vanishes by gauge invariance of the top Yukawa. We expect this cancellation occurs in any $U(1)_X$ extension of the SM that permits the top Yukawa. 

One interesting consequence of this is that, in such $U(1)_X$ extensions of the SM, a tree-level coupling to the Higgs is needed to obtain a sizeable shift in $M_W$, since large $C_{HD}$ will not be generated by RGE alone. This means the Higgs must be charged under $U(1)_X$, and therefore $U(1)_X$ must be chiral (at least, $q_L^3$ and $t_R$ are charged differently), as is the case for the $Z'$ models studied in {\em e.g.}~\cite{Allanach:2021kzj,Allanach:2022bik}. On the other hand, we do not expect vector-like extensions ({\em e.g.} from gauging $X=B_i - L_j$) to give sizeable corrections to $M_W$ from the gauge sector.

\subsection{Flavour-violating constraints}
As discussed in \S~\ref{sec:light-heavy-mixing}, flavour constraints combined with naturalness require that the CKM matrix comes dominantly from the down-quark sector. In the limit of up-alignment, the low-energy impact of the $U(2)^5$ breaking is completely fixed by the fermion masses and the CKM. In particular, flavour violation is present only in the left-handed down-quark sector and is completely determined by the CKM, such that we are effectively working in the MFV limit. As we will see, this leads to highly predictive phenomenology in $B$- and $K$-meson mixing, as well as in $B$-decays.

\subsubsection*{Meson mixing}
The low-energy effective Lagrangian describing meson mixing in the down-quark sector is
\begin{equation}
\mathcal{L}_{\Delta F = 2} = -C_{B_s}^1 (\bar s_L \gamma_\mu b_L)^2 - C_{B_d}^1 (\bar d_L \gamma_\mu b_L)^2 - C_{K}^1 (\bar d_L \gamma_\mu s_L)^2 \,.
\end{equation}
We have the MFV-like prediction where all the Wilson coefficients can be written in terms of $ C_{B_s}^1$,
\begin{equation}
 C_{B_s}^1 = \frac{X_q^2 }{2M_{Z'}^2} (V_{is}^* \, g_{Z'}^{ij} V_{jb})^2\,, \hspace{10mm} C_{B_d}^1 = \left(\frac{V_{td}^*}{V_{ts}^*}\right)^2 C_{B_s}^1 \,, \hspace{10mm} C_K^1 = \left(\frac{V_{td}^* V_{ts}}{V_{ts}^* V_{tb}}\right)^2 C_{B_s}^1  \,,
\end{equation}
such that all current bounds are automatically passed if the bound on $|C_{B_s}|$ is satisfied~\cite{Silvestrini:2018dos,EuropeanStrategyforParticlePhysicsPreparatoryGroup:2019qin}. As emphasized in~\cite{Panico:2016ull}, this relation between $B_s$, $B_d$, and kaon mixing could be tested experimentally in the future. Performing the sum and using CKM unitarity, we find the following exact expression
\begin{equation}
 C_{B_s}^1 = \frac{2X_q^2 g_Y^2}{M_{Z'}^2} \frac{(V_{ts}^* V_{tb})^2}{\sin^2 2\theta} = 1.2\times 10^{-5}~\text{TeV}^{-2} \left(\frac{\rm 1~ TeV}{M_{Z'}\sin 2\theta}\right)^2   \,,
\end{equation}
which should be compared to the bound $|C_{B_s}^1| < 2\times 10^{-5}~{\rm TeV}^{-2}$. We anticipated in the introduction that the bound on FCNC generated at tree-level but protected by MFV should be order 10 TeV without an additional suppression mechanism at work. While meson mixing is indeed induced at tree-level in the DH model, the suppression mechanism here comes from the smallness of both $g_Y$ and the quark doublet hypercharge $X_q$, leading to only a weak bound of $M_{Z'}\sin 2\theta \gtrsim 1$ TeV. Thus, minimally broken $U(2)$ (which allows only fully left-handed FCNC as in the SM) accidentally provides a higher degree of flavour protection in the DH model due to the chiral nature of the $Z'$ and the smallness of $X_q$, as was found in the `$Y_3$-like' models of~\cite{Allanach:2018lvl,Allanach:2019iiy,Davighi:2021oel}, which we find to be a particularly beautiful feature.\footnote{For comparison, in a model based on deconstructing $SU(2)_L$, one expects $B_{s,d}$ mixing to give a bound on the scale which is $|g_L T^3_d / (g_Y X_q)|\approx 5.4$ times stronger, forcing the model to less natural parameter space.}

\subsubsection*{$b\rightarrow s\ell\ell$ transitions}
\begin{figure}
    \centering
    \includegraphics[width=0.6\columnwidth]{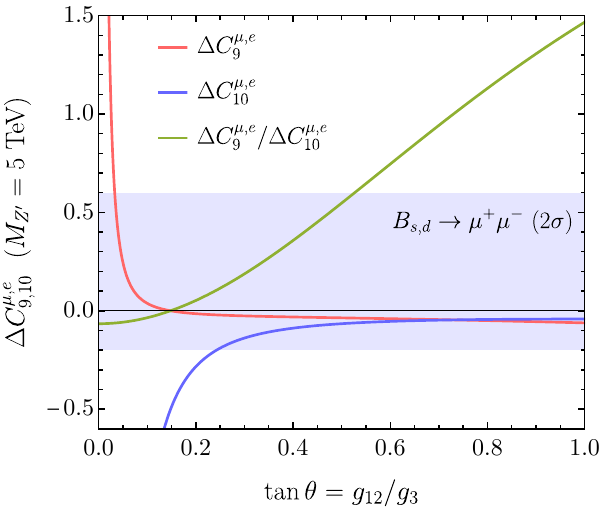}
    \caption{DH model predictions for $\Delta C_{9,10}^{\mu,e}$ and their ratio as a function of the gauge mixing angle, for $M_{Z'} = 5$ TeV. The WCs scale as $M_{Z'}^{-2}$, so the results can be easily rescaled to other $Z'$ mass values. The blue region gives the $\Delta C_{10}^{\mu}$ region preferred at $2\sigma$ by $\mathcal{B}(B_{s,d} \to \mu^+ \mu^-)$, which the blue curve must lie within, allowing one to also read off the corresponding prediction for $\Delta C_{9}^{\mu,e}$.}
    \label{fig:Cplot}
\end{figure}
New physics effects in the $b\to s\ell\ell$ system can be parameterized in the weak effective theory (WET), via the Lagrangian
\begin{equation}
    \mathcal{L}_{b\rightarrow s\ell\ell} = \frac{4G_F}{\sqrt{2}} V_{tb} V_{ts}^\ast \sum_i (C_i^{\rm SM} + \Delta C_i) \mathcal{O}_i\, ,
\end{equation}
where $\Delta C_i$ denotes the NP contribution to the WC for the operator $\mathcal{O}_i$. Of particular interest here are the operators
\begin{align}
    \mathcal{O}_9^{\ell} &= \frac{\alpha}{4\pi}(\overline{s} \gamma_\mu P_L b)(\overline{\ell}\gamma^\mu \ell)\,, \qquad \ell \in \{e,\mu\}\,, \\
    \mathcal{O}_{10}^{\ell} &= \frac{\alpha}{4\pi}(\overline{s} \gamma_\mu P_L b)(\overline{\ell}\gamma^\mu \gamma_5 \ell)\, ,
\end{align}
where $4\pi \alpha = e^2$.
We find the following expressions for the NP contributions that come from integrating out our $Z'$ at tree-level: 
\begin{align}
\Delta C_9^{\mu,e} &= \frac{\pi v^2}{\alpha}\frac{V_{is}^* V_{j b} }{V_{ts}^* V_{tb} }  [C_{lq}^{(1)}]_{22 i j} \left(1+\frac{X_e}{X_l} + \frac{X_H}{X_l}\frac{\zeta}{\tan^2\theta} \right)\,, \\
\Delta C_{10}^{\mu,e} &= \frac{\pi v^2}{\alpha}\frac{V_{is}^* V_{j b} }{V_{ts}^* V_{tb} }  [C_{lq}^{(1)}]_{22 i j} \left( -1 + \frac{X_e}{X_l} - \frac{X_H}{X_l}\frac{1}{\tan^2\theta} \right)\,,
\label{eq:C10}
\end{align}
where $\zeta = 1-4s_W^2$. The terms proportional to 1, $X_e$, and $X_H$ take into account the contributions to $\Delta C_{9,10}^{\mu,e}$ from $[C_{lq}^{(1)}]_{22 i j}$, $[C_{qe}]_{i j 22}$, and $[C_{Hq}^{(1)}]_{i j}$ respectively, where the SMEFT coefficients are given by the formulae in~\cref{eq:semi-leptonics}. For the ratio of WCs, we find 
\begin{align}
\frac{\Delta C_9^{\mu,e}}{\Delta C_{10}^{\mu,e}} = \frac{X_e + X_l + \zeta  \, X_H \cot^2 \theta}{X_e - X_l - X_H \cot^2 \theta}\,.
\end{align}
We plot the results for $\Delta C_9^{\mu,e}$, $\Delta C_{10}^{\mu,e}$, and their ratio in Fig.~\ref{fig:Cplot} as a function of $\tan\theta$, for $M_{Z'} = 5$ TeV. 

As the WCs are proportional to $M_{Z'}^{-2}$, rescaling the result to other values of the $Z'$ mass is straightfoward. Because of $e-\mu$ universality, there is no NP contribution  to the LFUV ratios $R_{K^{(\ast)}}$ (in accordance with the recent measurements from LHCb~\cite{LHCb:2022qnv}). There is, on the other hand, a significant NP effect in the $B_{s,d} \to \mu^+ \mu^-$ branching ratio, due to a large $\Delta C_{10}$ contribution that increases as $\tan\theta$ decreases, as can be seen in~\cref{eq:C10} and \cref{fig:Cplot}. This is due to the fact that, while the direct coupling of the $Z'$ to muons is reduced, the mixing with the SM $Z$ increases, captured in the SMEFT by $[C_{Hq}^{(1)}]_{23} \propto X_H \cot^2\theta$. This $Z-Z'$ mixing causes the $Z'$ to inherit the accidentally small vector coupling $\zeta$ of the SM $Z$ to charged leptons, resulting in an almost pure $ C_{10}$ contribution. The theoretical expression for $\mathcal{B}(B_{i} \to \mu^+ \mu^-)$ reads
\begin{equation}
\frac{\mathcal{B}(B_{i} \to \mu^+ \mu^-)}{\mathcal{B}(B_{i} \to \mu^+ \mu^-)_{\rm SM}} = \bigg|1+\frac{[\Delta C_{10}^{\mu}]_i}{C_{10}^{\rm SM}} \bigg|^2 \,,
\end{equation}
where $[\Delta C_{10}^{\mu}]_s = \Delta C_{10}^{\mu}$, $[\Delta C_{10}^{\mu}]_d = \Delta C_{10}^{\mu} (s\rightarrow d)$, and $C_{10}^{\rm SM} = -4.2$. Since $\mathcal{B}(B_{s,d} \to \mu^+ \mu^-)$ are correlated observables, we take these constraints into account using the combined likelihood given in Ref.~\cite{Greljo:2022jac}, which also includes an up-to-date average of all the experimental measurements. We use this likelihood to obtain the $2\sigma$ preferred region for $\mathcal{B}(B_{s,d} \to \mu^+ \mu^-)$ shown in Fig.~\ref{fig:Cplot}, which the blue curve must lie within.

As a final comment, we note that our DH model predicts $\Delta C_{L,\nu}^{\mu,e} = \Delta C_{10}^{\mu,e}$ (see {\em e.g.}~\cite{Altmannshofer:2009ma,Belle-II:2018jsg} for the definition of $C_{L,\nu}$) which means that the bound from $\mathcal{B}(B_{s} \to \mu^+ \mu^-)$ prevents any large enhancement (beyond a few percent) to $\mathcal{B}(B \rightarrow K \nu \bar\nu)$. The crucial point, which is worth re-emphasizing, is that the $Z-Z'$ mixing does not allow one to decouple effects in the light leptons by going to the small gauge mixing angle limit. Furthermore, considering third-generation leptons specifically, the model predicts $\Delta C_{L,\nu}^{\tau} = \Delta C_{10}^{\tau} = 0$. Explicitly, we find
\begin{equation}
\Delta C_{L,\nu}^{\tau} = \Delta C_{10}^{\tau} \propto  X_H -X_l + X_e  = 0\,,
\label{eq:CLtau}
\end{equation}
which means that our model gives no contribution to $\mathcal{B}(B_{s} \to \tau^+ \tau^-)$ or $\mathcal{B}(B \rightarrow K \nu \bar\nu)$ involving $\tau$-neutrinos. We note that the relation between charges giving the cancellation follows from the gauge invariance of the tau Yukawa coupling in our model.

\subsection{Combined results and future projections}
\label{sec:combResults}
\begin{figure}
    \centering
    \hspace{-5mm}
    \includegraphics[width=0.485\columnwidth]{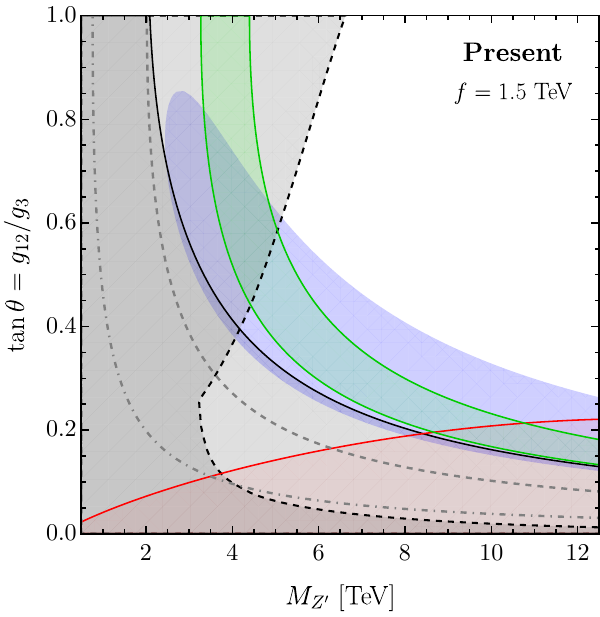} \hspace{5mm}
    \includegraphics[width=0.485\columnwidth]{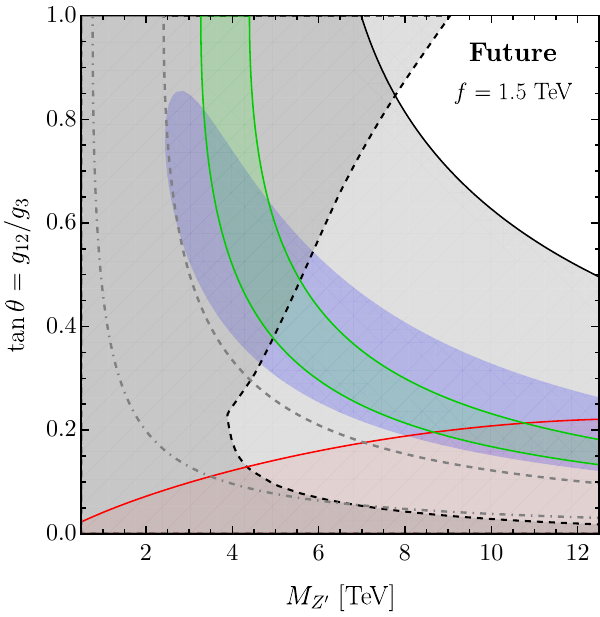}
    \caption{Blue gives the 1$\sigma$ region preferred by the EW fit using $m_W^{\rm new}$, while the solid black line gives the 2$\sigma$ exclusion from the EW fit using $m_W^{\rm old}$ average (\S~\ref{sec:EWPO}). The green band is the region where a satisfactory explanation of the $m_c/m_t$  mass hierarchy is achieved (\S~\ref{sec:lightYuks}). The dashed black line gives the region excluded by $pp\rightarrow ee,\mu\mu, \tau\tau$ searches for the $Z'$ at high-$p_T$. The dashed gray line gives the region excluded by $B_{s,d} \rightarrow \mu^{+}\mu^{-}$, while dot-dashed gray line gives the region excluded by meson mixing.  Finally, the region below the solid red line is where radiative Higgs mass corrections from the $Z'$ exceed $\delta M_{H_l}^2 > f^2$. The left plot shows current constraints, while the right plot shows projections for the future including 1) a HL-LHC projection with $3~\text{ab}^{-1}$ (black dashed), 2) a projection for $B_{s,d} \rightarrow \mu\mu$ from LHCb, and 3) an FCC-ee projection for the region excluded by EWPO (solid black). See \S~\ref{sec:combResults} for further details.
    }
    \label{fig:moneyPlot}
\end{figure}
Having discussed all flavour conserving, flavour violating, and naturalness constraints, we are now ready to see how the $\tan\theta$ vs $M_{Z'}$ parameter space of the DH model is globally constrained. To this end, the left panel of~\cref{fig:moneyPlot} summarizes the current $95\%$ CL constraints in the $\tan\theta$ vs $M_{Z'}$ plane, for $f=1.5$ TeV. In particular, we plot the flavour-conserving constraints from high-$p_T$ (black dashed) and EWPOs (black) and flavour-violating constraints from meson mixing (gray dot-dashed) and $B_{s,d}\rightarrow \mu^+ \mu^-$ (gray dashed). The red shaded region below the solid red line gives the region excluded by finite naturalness, {\em i.e.} the region where $\delta {M^2_{H_l}} > f^2$ due to radiative Higgs mass corrections from the $Z'$.  The change in behavior of the high-$p_T$ bound around $(M_{Z'} = 3.5~\text{TeV}, \tan\theta =0.25$) is where the bound from the $pp\rightarrow \tau\tau$ ATLAS search becomes stronger than the $pp \rightarrow \ell\ell $ search from CMS. We see that the parameter space is dominantly constrained by the flavour-conserving bounds: a combination of high-$p_T$ searches and EWPO require $M_{Z'} \gtrsim 4.5$ TeV. Therefore, the viable parameter space of the model does not depend on the choice of up- or down-alignment. We use $M_{Z'} \gtrsim 4.5$ TeV to obtain a lower bound on the overall scale of the model $f\gtrsim 1.3$ TeV for $\vh = f$, $\vq = 4\pi f$, and $\tan\theta = 0.4$. As a final point we note that naturalness together with EW bounds require $\tan\theta \gtrsim 0.2$. Using the 2-loop running available in the \texttt{RGBeta} package~\cite{Thomsen:2021ncy}, we find this translates into the Landau pole for $U(1)_3$ appearing at a scale $\Lambda_{g_3 = 4\pi} \gtrsim 10^4 \times f$.

Next, we discuss the parameter space preferred by the model. In particular, the shaded blue gives the $1\sigma$ region preferred by the EW fit using $M_W^{\rm new} = 80.410(15)$ GeV. As discussed in detail in (\S~\ref{sec:EWPO}), this region is not excluded by the EW fit using the old value of $M_W^{\rm old} = 80.379(12)$ GeV (solid black line). This is because the EW exclusion bound is dominated by $Z\rightarrow e_R \, e_R$ for $\tan\theta \approx 1$, where it requires $M_{Z'} \gtrsim 2$ TeV, becoming stronger as $\tan\theta$ decreases, where the bound comes mainly from $Z\rightarrow \tau_R\,  \tau_R$. The overall conclusion is that $C_{HD}$, which is directly related to the EW $\rho$-parameter, does not ever provide the dominant constraint from EWPO. Therefore, some NP effect in $C_{HD}$ is preferred for $M_W^{\rm new}$, asking the model to live in the blue-shaded region where the $Z'$ is relatively light. Interestingly, this region has a strong overlap with the shaded green region where the mass hierarchy $m_c \ll m_t$ is naturally explained (see~\S~(\ref{sec:lightYuks}) for details). This is one key result of our analysis in this work: \emph{the natural parameter space of the DH model (given by the green band) necessarily implies positive large shifts in $M_W$ as well as sizeable enhancements to $\mathcal{B}(B_{s,d} \to \mu^+ \mu^-)$.}

 The right panel of~\cref{fig:moneyPlot} gives projections for how constraints on the DH model parameter space are expected to improve in the future, with more data from LHCb, HL-LHC, and a potential future EW precision machine such as FCC-ee. In particular, the changes with respect to the left panel are: 
 \begin{enumerate}
 \item An HL-LHC projection for the $pp\to \ell\ell/\tau\tau$ search constraints at high-$p_T$, which we calculate assuming an integrated luminosity of $3~\text{ab}^{-1}$ and that the bound on the cross section improves as $\sqrt{L}$ (black dashed); 
 \item A future projection for the bound coming from $\mathcal{B}(B_{s,d} \to \mu^+ \mu^-)$, where we assume LHCb measures the central value of the SM theory predictions with uncertainties given in~\cite{LHCb:2018roe} (gray dashed); 
 \item An FCC-ee projection for the constraint coming from the electroweak fit; we assume the statistical error on all EW precision observables is reduced by a factor 10, while the central values are given by the SM theory predictions (solid black).
 We also assume that we are not limited by systematic or theory uncertainties. This precision should be achieved for $2\times 10^{11}$ Z-bosons ($10^4$ more than LEP) out of the planned $5\times 10^{12}$ Z-bosons, or 1/25 of the total $Z$ events, which should occur in the first 3 months of data taking~\cite{Bernardi:2022hny}. 
  \end{enumerate}
In summary, we see that FCC-ee is the best machine to probe the DH model, as it easily has the reach to fully exclude the entire natural parameter space. We expect that this is a generic conclusion for flavour models that explain both fermion masses and mixings at a low scale and/or models solving the EW hierarchy problem, due to the requirement of direct Higgs couplings leading to large effects in EWPO.

\section{Concluding Remarks}
\label{sec:discussion}

As the LHC program continues to mature, it remains strongly motivated to thoroughly explore the landscape of viable low-scale ({\em i.e.} TeV) flavour models, and to constrain them via the two-pronged strategy of high-$p_T$ searches at the LHC and increasingly precise measurements of rare heavy-flavoured decays that test the accidental approximate $U(2)^5$ symmetry of the SM. In this paper we have introduced an economical (but complete) low-scale flavour model, based on deconstructing the SM hypercharge gauge symmetry, and elucidated its phenomenology for colliders, EW precision, and flavour observables. Like any such flavour-deconstructed gauge model seeking to explain fermion masses and mixings, there are heavy gauge bosons (here a $Z'$) that couple directly to the Higgs. This introduces an interesting interplay where high-$p_T$ bounds can be avoided by reducing the coupling to light fermions, but only at the expense of increased couplings to the third family and the Higgs, leading to conflict with EW precision data. We find that the combination of EWPO together with high-$p_T$ searches require $M_{Z'} \gtrsim 4.5$ TeV in our DH model.
By computing 1-loop Higgs mass corrections, we find the model is nonetheless natural up to an inevitable fine-tuning of $v^2/f^2 \sim 1\%$, corresponding to the usual little hierarchy problem- now an experimentally established fact. However, the remaining large hierarchy problem can still be resolved by embedding our model in a supersymmetric or composite theory at a higher scale $4 \pi f \sim 12$ TeV.

We now comment on possibilities for further NP at yet higher scales $\gtrsim 4\pi f$. Three puzzles that could be explained deeper in the UV are (1) the quantization of hypercharge, (2) the origin of the light-family mass hierarchy $m_1 \ll m_2$, and (3) the origin of neutrino masses. 
Puzzle (1) could be solved by embedding DH inside a semi-simple gauge group, which interestingly requires flavour non-universal quark-lepton unification. If we discount options that violate $B-L$, there are essentially two options that preserve universal $SU(2)_L$ symmetry. Both take the form $SU(2)_L \times [K]_{12} \times [SU(4)\times SU(2)_R]_3$, where either $[K]_{12}=[SU(4)\times Sp(4)_R]_{12}$, unifying flavour with $SU(2)_R$, or $[K]_{12}=[SU(8) \times SU(2)_R]_{12}$, unifying flavour with colour. To also explain puzzle (2), the former of these options is preferred, offering a purely-RH version of `electroweak flavour unification' ~\cite{Davighi:2022fer,Davighi:2022bqf}, which, in the 1-2 sector, predicts $m_1 \ll m_2$ and an order-1 Cabibbo angle~\cite{Davighi:2023iks}. If one only explains puzzle (2) and postpones the quantization of hypercharge, then the obvious UV completion is via hypercharge cubed symmetry~\cite{FernandezNavarro:2023rhv}. 
Puzzle (3), the origin of neutrino masses, could be solved in the deep UV via a multi-scale form of the usual see-saw mechanism (See Appendix~\ref{app:neutrino} for some discussion.) The crucial point is that the electroweak and $f$ scales are protected against all of these `deep UV' layers of NP if there is SUSY or compositeness at the scale $4\pi f$.

A particularly compelling scenario (in which all of the above can be implemented) is where the minimally broken $U(2)^5$ symmetry studied here emerges dynamically from a multi-scale theory of flavour~\cite{Panico:2016ull,Bordone:2017bld,Fuentes-Martin:2020pww,Fuentes-Martin:2022xnb}, with one family of SM fermions and a Higgs field for each scale (or site). Minimally-broken $U(2)$ is then realized if the left-handed fermions and Higgs fields mix between sites, while right-handed fermions are fully localised. In this picture, the third-family scale where $H_3$ is localised should lie as low as possible to minimise fine-tuning. This prefers scenarios such as our DH model which can be realized safely at the TeV scale, even when including third-family quark-lepton unification~\cite{Greljo:2018tuh,Allwicher:2023aql}.\footnote{In particular, a straightforward $SU(4)_3 \times SU(3)_{12} \times SU(2)_L \times U(1)_R \times U(1)_{12}$ extension of DH could also address the charged-current $B$-meson anomalies~\cite{Greljo:2018tuh,Aebischer:2022oqe}, which are perhaps hinting at low-scale quark-lepton unification. In particular, the new coloured gauge bosons do not give Higgs mass corrections at 1-loop~\cite{Davighi:2023iks}, and therefore do not change our conclusions regarding naturalness.} We emphasize that this multi-scale setup is an inversion of the prevailing view before the LHC, which focused on first solving the large hierarchy problem at the TeV scale, while postponing a solution to the flavour puzzle to higher scales.

Following this logic, we have proposed a scenario where the flavour puzzle is first partially solved at a low scale of $f \sim 1$ TeV, by particular gauge dynamics (deconstructed hypercharge) that (i) remains phenomenologically viable at this low scale, and (ii) gives at most TeV scale radiative contributions to the Higgs mass, thereby not worsening the little hierarchy problem. We then entertain the possibility that the large hierarchy problem is resolved at a higher scale of $4\pi f \sim 12$ TeV, consistent with taking current experimental bounds at face value. The leading low-energy phenomenology, due to a single $Z'$ gauge boson, includes large positive shifts in the $W$-boson mass, as well as an enhancement in $\mathcal{B}(B_{s,d} \to \mu^+ \mu^-)$. 

Our DH $Z'$ gauge boson can be searched for at the LHC, where we have emphasized that a combined strategy involving all di-lepton final states is necessary, due to the $U(2)^5$ symmetry of the gauge sector. In particular, searches in di-tau final states should also include a $b$-tag category, since the $Z'$ is dominantly produced via $bb\rightarrow \tau\tau$ in the small mixing angle limit. To help facilitate such searches, we have uploaded a UFO model file to the \texttt{arXiv} which implements the DH model for use in experimental studies. We hope that this model serves as a benchmark study for the more general idea of $U(2)$ based NP searches, which in particular allow for the well-motivated option of NP coupled dominantly to the third family that is less constrained by current search strategies.

Finally, as we have demonstrated here (and echoing the pre-LHC message advocated in Ref.~\cite{Barbieri:2000gf}), a future electroweak precision machine like the FCC-ee is the best way to probe NP with sizeable couplings to the Higgs, as always occurs in low-scale flavour models and/or models solving the EW hierarchy problem. We therefore conclude by stating our belief that the next European strategy meeting should finalize a plan to build FCC-ee.

\section*{Acknowledgments}

We thank Ben Allanach, Admir Greljo, Gino Isidori, Matthew Kirk, Javier M. Lizana, Patrick Owen, and Felix Wilsch for useful discussions. We also thank Felix Wilsch in particular for his guidance in extracting the collider bounds on the $Z'$. 
JD is grateful to Ben Allanach for collaboration on related projects exploring the phenomenology of third family hypercharge model(s).

\appendix

\section{EFT for neutrino masses}
\label{app:neutrino}

For completeness, we here sketch a possible setup for realising neutrino masses in our model, which we phrase in terms of an EFT description including both $H_3$ and $H_{12}$ Higgs fields. 
We note that Ref.~\cite{FernandezNavarro:2023rhv} suggests a more detailed account, using a variety of auxiliary fields, of the origin of specific PMNS matrix elements in the related `tri-hypercharge' model. We begin by observing that the DH EFT allows for Weinberg operators of the type
\begin{align}
&\mathcal{L}_{d=5} \supset \frac{C_{R}^{33}}{\Lambda_{R,h}} (\bar\ell_L^{3} \tilde{H}_3)(\tilde{H}_3^T \ell_L^{c3}) + \frac{C_{R}^{i3}}{\Lambda_{R,hl}} (\bar\ell_L^{i} \tilde{H}_{12})(\tilde{H}_3^T \ell_L^{c3}) + \frac{C_{R}^{ij}}{\Lambda_{R,l}} (\bar\ell_L^{i} \tilde{H}_{12})(\tilde{H}_{12}^T \ell_L^{cj}) + {\rm h.c.} \,
\end{align}
Assuming $O(1)$ WCs for these operators, the rough condition for an anarchic neutrino mass matrix is then
\begin{equation}
\frac{y_c^2 v^2}{\Lambda_{R,l}}\sim \frac{y_c v^2}{\Lambda_{R,hl}} \sim \frac{v^2}{\Lambda_{R,h}}   \sim 0.05~{\rm eV} \,,
\end{equation}
which points to a scale hierarchy of $(\Lambda_{R,l},\Lambda_{R,hl}, \Lambda_{R,h}) \sim (10^{11}, 10^{13}, 10^{15})$ GeV.  Thus, the neutrino mass puzzle should be solved inversely, {\em i.e.} for the light generations first at the scale $\Lambda_{R,l} \sim 10^{11}$ GeV ($10^{7}$ GeV if we interpolate to the first family) and finally for the third generation at the GUT scale $\Lambda_{R,h} \sim 10^{15}$ GeV. A simple UV origin for these scales would be to extend the model by three RH neutrinos, which are SM singlets, and where the scales $\Lambda_{R,l}$, $\Lambda_{R,hl}$ and $\Lambda_{R,h}$ appear in front of (effective) Majorana mass terms. We re-iterate that, if supersymmetry or compositeness were manifest at the scale $4\pi f \ll 10^7$ GeV, then the Higgs mass parameter would not receive loop corrections scaling quadratically with these high scales (but could depend on them logarithmically).

In our conclusions, we discuss the possibility of third-family quark-lepton unification via $SU(4)_3$ (in order to explain, for example, the quantization of hypercharge). By putting the $\nu_R^3$ field in the same multiplet as $t_R$, a Majorana mass is not permitted in the $SU(4)_3$ invariant phase and so the scale $\Lambda_{R,h}$ would then be tied to the scale of $SU(4)_3$ breaking. 
In this scenario, the high scale seesaw scheme proposed above is only consistent when $SU(4)_3$ is broken above the GUT scale. For a lower scale $SU(4)_3$ breaking,
one has to implement the inverse seesaw (ISS) mechanism, which would give effective dimension-5 Weinberg operators scaling as $\mu/\Lambda_{R,h}^2$, where $\mu$ is a (technically natural) small lepton number breaking parameter. This works even for the $SU(4)_3$ breaking scale as low as $\sim$ TeV, as shown in~\cite{Greljo:2018tuh,Fuentes-Martin:2020pww}.

\bibliographystyle{JHEP}
\bibliography{refs}

\end{document}